\documentclass{article}
\usepackage{amssymb}

\usepackage{graphicx}
\usepackage{amsmath}

\input{tcilatex}

\begin{document}

\title{\textbf{High-}$\mathbf{T}_{c}$\textbf{\ Superconductivity via BCS and
BEC Unification: A Review}}
\author{M. de Llano \and Instituto de Investigaciones en Materiales, UNAM,
04510 M\'{e}xico, DF, Mexico}
\maketitle

\begin{abstract}
\ \ Efforts to unify the Bardeen, Cooper \& Schrieffer (BCS) and the
Bose-Einstein condensation (BEC) formalisms in terms of a ``\textit{complete}
boson-fermion (BF) model'' (CBFM) are surveyed. A vital distinction is that
Cooper pairs (CPs) are indeed bosons that suffer BEC, in contrast with BCS
pairs that are not bosons. Another crucial ingredient (particularly in 2D
where ordinary BEC does not occur) is the \textit{linear }dispersion
relation of ``ordinary'' CPs, at least in leading order in the
center-of-mass momentum (CMM) power-series expansion\ of the CP energy. This
arises because CPs propagate not \textit{in vacuo }but in the Fermi ``sea.''
A many-body Bethe-Salpeter equation treatment of CPs based on the ideal
Fermi gas (IFG) sea\ yields the familiar negative-energy, two-particle
bound-state \textit{if} 2h-CPs are ignored as in the ordinary CP problem.
But it gives purely-imaginary energies, and is thus meaningless, if 2h-CPS
are included as completness requires. However, when based on the BCS ground
state instead of the IFG, in addition to the familiar trivial solution (or
Anderson-Bogoliubov-Higgs) sound mode, legitimate two-particle \textit{moving%
}\ ``generalized CPs'' emerge but as positive-energy, finite-lifetime,
resonant nontrivial solutions for nonzero-CMM. This amounts to replacing the
purely-kinetic-energy unperturbed Hamiltonian by the BCS one. The moving CPs
again have a \textit{linear }dispersion leading term. BEC of such pairs may
thus occur in exactly 2D (as it cannot with quadratic dispersion) and in
fact all the way down to ($1+\epsilon $)D where $\epsilon $ can be
infinitesimally small, thus encompassing all empirically known
superconductors.\ 

\ \ \ \ The unified theory reduces in limiting cases to all the main
continuum (as opposed to ``spin'') statistical theories of
superconductivity. These include both the BCS and BEC theories. The unified
BF theory is ``complete'' in that not only{\large \ }two-electron (2e) but
also two-hole (2h) CPs are allowed, and in arbitrary proportions. In
contrast, BCS theory can be deduced from the CBFM but allows only equal
(50\%-50\%) mixtures of them, a fact rarely if ever stressed. The CBFM shows
that the BCS condensate is precisely a BE condensate of a mixture of
kinematically independent electrons coexisting with weakly-coupled zero CMM
2e- and 2h-CPs in \textit{equal} proportions. Without abandoning the
electron-phonon mechanism, the CBFM has been applied in 2D and 3D. The BCS
model interaction in moderately weak coupling is sufficient to reproduce the
unusually high values of $T_{c}$ (in units of the Fermi temperature) of $%
0.01-0.1$\ empirically exhibited by the so-called ``exotic''
superconductors, including cuprates. This range is high relative to the low
values of $\leq 10^{-3}$ more or less correctly reproduced by BCS theory for
conventional (mostly elemental) superconductors. Also accounted for is the
empirical fact that ``hole superconductors'' systematically have higher $%
T_{c}$'s. Room temperature superconductors are predicted to be possible but
only via 2h-CP BE condensates.

\ \ \ \ 

\textit{Running Title}: \textbf{BCS and BEC Unification: A Review}
\end{abstract}

\section{Introduction}

Boson-fermion (BF) models of superconductivity (SC) as a Bose-Einstein
condensation (BEC) \cite{Ogg}\cite{Ginz}\ go back to the mid-1950's \cite%
{Blatt}-\cite{BF2}, pre-dating even the BCS-Bogoliubov theory \cite{bcs}-%
\cite{bts}. Although BCS theory only contemplates the presence of ``Cooper
correlations'' of single-particle states, BF models 
\cite{Blatt}-\cite{BF2},\cite{BF3}-\cite{BF10} posit the existence of actual
bosonic CPs. Such pair charge carriers have been observed in magnetic flux
quantization experiments on elemental \cite{classical,classical2} as well as
on cuprate \cite{cuprates}\ superconductors (SCs). Larger clusters than
pairs are \textit{not }observed, apparently because the clustering occurs
not \textit{in vacuo }but in the Fermi sea. However, no experiment has yet
been done, to our knowledge, that{\large \ }distinguishes between electron
and hole CPs. CPs appear to be the single most important universally
accepted ingredient of SC, whether conventional or ``exotic'' and whether of
low- or high-transition-temperatures $T_{c}$. And yet, inspite of their
centrality they are poorly understood. The fundamental drawback of early %
\cite{Blatt}-\cite{BF2} BF models, which took 2e bosons as analogous to
diatomic molecules in a classical atom-molecule gas mixture, is the
notorious absence of an electron energy gap $\Delta (T)$. ``Gapless'' models
cannot describe the superconducting state at all, although they are useful
in locating transition temperatures if approached from above, i.e., $T>T_{c}$%
. Even so, we are not aware of any calculations with the early BF models
attempting to reproduce any empirical $T_{c}$ values. The gap first began to
appear in later BF models \cite{BF3}-\cite{CMT02}. With two \cite{BF7a}\cite%
{PLA2}\ exceptions, however, all BF models neglect the effect of \textit{%
hole }CPs included on an equal footing with electron CPs to give\ a
``complete'' BF model (CBFM) consisting of \textit{both} bosonic CP species
coexisting with unpaired electrons.

For 13 years the highest $T_{c}$ value for any superconductor was $23$ $K$,
until the discovery \cite{BM86}\ in 1986 of the first so-called ``high-$%
T_{c} $'' cuprate superconductor $LaBaCuO$ having a $T_{c}\simeq 35$ $K$. A
feverish search for materials with even higher $T_{c}$'s lead within just
seven years to the highest-$T_{c}$ superconductor known to date, the $%
HgBaCaCuO$ cuprate \cite{Chu93} with a $T_{c}\simeq 164$ $K$ under very high
pressure ($\simeq 310,000$ atm). The embarrassing fact is that since 1993
this record has not been broken, very probably because there is yet no 
\textit{predictive} microscopic theory of superconductivity that can provide
orientation in a search that until now has proceeded on a trial-and-error
basis.

We submit that progress in developing such a theory has been held back by
too many common myths or false dogmas firmly entrenched in the theoretical
community.

\section{Some false dogmas}

The false theoretical dogmas just mentioned can be summarized in the
following assertions:

1. With the electron-phonon dynamical mechanism transition temperatures (as
given by the BCS formula) $T_{c}=1.13\Theta _{D}e^{-1/\lambda }\lesssim 45$ $%
K$ at most, since a typical Debye temperature $\Theta _{D}\sim 300K$ and $%
\lambda \lesssim 
{\frac12}%
$. For higher $T_{c}$'s\ one needs to invoke magnons or excitons or\
plasmons or other electronic mechanisms to provide pairing \cite{Hirsch02}.

2. Superconductivity is unrelated to Bose-Einstein condensation (BEC) \cite%
{Bardeen}.

3. BEC is impossible in 2D \cite{AAA}\cite{ASA}.

4.\ Cooper pairs: \ 

\qquad a\textbf{) }are such that \textbf{``}there is a very strong
preference for singlet, zero-momentum pairs, so strong that one can get an
adequate description of SC by treating these correlations alone \cite{Coo63}%
.''

\qquad b) consist of negative-energy stable (i.e., stationary) bound states %
\cite{Coo}.

\qquad c) propagate in the Fermi sea with an energy $\hbar ^{2}K^{2}/2(2m)$,
where $m$ is the effective electron (or hole) mass and $\hbar K$ the CP
center-of-mass momentum (CMM) \cite{PRLreferee}, hence assertion \# 3 above.

\qquad d) with the linear dispersion $E$\ $\varpropto $\ $v_{F}\hbar K$,\
where $v_{F}$ is the Fermi velocity, simply represent the sound mode of the
ideal Fermi gas (+ interactions), with sound speed $v_{F}/\sqrt{d}$ in any
dimensionality $d$ \cite{Randeria96}.

\qquad e) ``...with $K\neq 0$\ represent states with net current flow'' \cite%
{NobelLecture}.

\qquad f) and BCS pairs are the same thing \cite{Clem03}.

\qquad g) are \textit{not} bosons, Ref. \cite{Schrieffer} p. 38, hence
assertion \# 2 above.

\section{Ordinary Cooper pairing}

For bosons with excitation (e.g., kinetic)\ energy for small CMM $K$ given
by 
\begin{equation}
\varepsilon _{K}=C_{s}K^{s}+o(K^{s}),  \label{0}
\end{equation}%
\ with $C_{s}$ some coefficient and $s>0$,\ BEC occurs in a box of length $L$%
\ if and only if $d>s$, since $T_{c}\equiv 0$ for all $d\leq s$. The
commonest example is $s=2$ as in the textbook case of ordinary bosons with
exactly $\varepsilon _{K}=$ $\hbar ^{2}K^{2}/2m_{B}$, with $m_{B}$ the boson
mass, giving the familiar result that BEC is not allowed for $d\leq 2$. The
general theorem for any $s>0$ is stated as follows. The total boson number
is 
\begin{equation}
N=N_{0}(T)+\sum_{\mathbf{K\neq 0}}[\exp \beta (\varepsilon _{K}-\mu
_{B})-1]^{-1}  \label{1}
\end{equation}%
with $\beta \equiv k_{B}T$. Since $N_{0}(T_{c})\simeq 0$ while the boson
chemical potential $\mu _{B}$ also vanishes\ at $T=T_{c}$, in the
thermodynamic limit the (finite) boson number density becomes%
\begin{equation}
N/L^{d}\simeq A_{d}\int_{0^{+}}^{\infty }\mathrm{d}KK^{d-1}[\exp \beta
_{c}(C_{s}K^{s}+\cdots )-1]^{-1}  \label{2}
\end{equation}%
where $A_{d}$ is a finite coefficient. Thus%
\begin{equation}
N/L^{d}\simeq A_{d}(k_{B}T_{c}/C_{s})\int_{0^{+}}^{K_{\max }}\mathrm{d}%
KK^{d-s-1}+\int_{K_{\max }}^{\infty }\cdots ,  \label{3}
\end{equation}%
where $K_{\max }$ is small and can be picked arbitrarily so long as the
integral $\int_{K_{\max }}^{\infty }\cdots $ is finite, as is $N/L^{d}$.
However, if $d=s$ the first integral\ gives $\ln K\mid _{_{0}}^{K_{\max
}}=-\infty $; and if $d<s$ it gives $1/(d-s)K^{s-d}\mid _{_{0}}^{K_{\max
}}=-\infty $. Hence, $T_{c}${\small \ }must vanish{\small \ }if and only if%
{\small \ }$d\leq s$,{\small \ }but is otherwise finite. This conclusion
hinges \textit{only }on the leading term of the boson dispersion relation $%
\varepsilon _{K}$.\ 

The case $s=1$ emerges in both the ``ordinary'' CP problem \cite{Coo}\ to be
recalled now, as well as in the ``generalized'' case of the next section.
The question of whether or not CPs are bosons or not is resolved in Appendix
A. For $s=1$ BEC occurs for all $d>1$.\ Striking experimental confirmation
of how superconductivity is ``extinguished'' as dimensionality $d$ is
diminished towards unity has been reported by Tinkham and co-workers \cite%
{Tinkham01}\cite{Tinkham00}. They report conductance\textit{\ vs.} diameter
curves in superconducting nanowires consisting of carbon nanotubes sputtered
with amorphous $Mo_{79}Ge_{21}$ ($T_{c}\simeq 5.5$ $K$) and of widths from
22 to 10 nm. The conductance diminishes to zero, implying the nanotube
becomes an insulator below a certain diameter, thus exposing how $T_{c}$
vanishes for the thinnest nanotubes.

The CP equation for the energy ${\mathcal{E}_{K}}$\ of two fermions above
the Fermi surface with momentum wavevectors $\mathbf{k}_{1}$ and $\mathbf{k}%
_{2}$ (and arbitrary CMM wavenumber $K$ where$\ \mathbf{K}\equiv \mathbf{k}%
_{1}+\mathbf{k}_{2}$) is given by 
\begin{equation}
\lbrack {\hbar ^{2}k^{2}/m-2E_{F}-\mathcal{E}_{K}+\hbar ^{2}K^{2}/}4m]\psi _{%
\mathbf{k}}=-\sum_{\mathbf{q}}{}^{^{\prime }}V_{\mathbf{kq}}\psi _{\mathbf{q}%
},  \label{4}
\end{equation}%
where $\mathbf{k}\equiv \frac{1}{2}(\mathbf{k}_{1}-\mathbf{k}_{2})$ is the
CP relative momentum and $\psi _{\mathbf{q}}$ its wave function in momentum
space. The prime on the summation implies restriction to states \textit{above%
} the Fermi surface with energy $E_{F}\equiv \hbar ^{2}k_{F}^{2}/2m$, viz., $%
|\mathbf{k}\pm \mathbf{K}/2|>k_{F}$, and $V_{\mathbf{kq}}$ is the double
Fourier transform of the interaction defined as 
\begin{equation}
V_{\mathbf{kq}}\equiv \frac{1}{L^{d}}\int d\mathbf{r}\int d\mathbf{r}%
^{\prime }e^{-i\mathbf{q}\cdot \mathbf{r}}V(\mathbf{r},\mathbf{r}^{\prime
})e^{i\mathbf{k}\cdot \mathbf{r}^{\prime }},  \label{4a}
\end{equation}%
with $V(\mathbf{r},\mathbf{r}^{\prime })$ the (possibly nonlocal)
interaction in real $d$-dimensional space.

\subsection{Delta interaction}

If the interfermion\ interaction $V(\mathbf{r},\mathbf{r}^{\prime })$ is
local, then $V(\mathbf{r},\mathbf{r}^{\prime })=V(\mathbf{r})\delta (\mathbf{%
r}-\mathbf{r}^{\prime })$ in (\ref{4a}). Moreover, if $V(r)=-v_{0}\delta (%
\mathbf{r})$ with $v_{0}>0$, this gives $V_{\mathbf{kq}}=v_{0}/L^{d}$ and (%
\ref{4}) becomes, for any $d$, 
\begin{equation}
\frac{1}{L^{d}}\sum_{\mathbf{k}}{}^{^{\prime }}\frac{1}{\hbar
^{2}k^{2}/m-2E_{F}-\mathcal{E}_{K}+\hbar ^{2}K^{2}/4m}=\frac{1}{v_{0}}.
\label{5}
\end{equation}%
In $d=1$ where the $\delta $-well supports a single bound state, the problem
is quite tractable \cite{Casas91}. In either 2D or 3D, however, the $\delta $%
-well supports an infinite set of bound levels with the lowest level in each
case being infinitely bound. This in turn leads to a rigorous collapse of
the many-fermion system \cite{AJPAS03}. To prevent this unphysical collapse
the $\delta $-wells must be ``regularized,'' i.e., constructed, say, from
square wells \cite{GT} such that the remaining $\delta $-well possesses only
one bound level. This leaves an infinitesimally small strength parameter $%
v_{0}$ which would make the rhs of (\ref{5}) diverge (so as to cancel the
lhs that also diverges in 2D and 3D but not in 1D). Combining (\ref{5}) for
2D with the vacuum two-body momentum-space Schr\"{o}dinger equation for the
same $\delta $-potential well allows eliminating \cite{PRB2000} $v_{0}$ in
favor of the (positive) binding energy $B_{2}$ of the single bound level of
the regularized $\delta $-well. One arrives at 
\begin{equation}
\sum_{\mathbf{k}}\frac{1}{B_{2}+\hbar ^{2}k^{2}/m}=\sum_{\mathbf{k}%
}{}^{^{\prime }}\frac{1}{\hbar ^{2}k^{2}/m-2E_{F}-\mathcal{E}_{K}+\hbar
^{2}K^{2}/4m},  \label{6}
\end{equation}%
where $B_{2}\geq 0$ now serves as a coupling constant. A small-$K$
power-series expansion for $\mathcal{E}_{K}$ gives the analytic expression 
\textit{valid for any dimensionless coupling }$B_{2}/E_{F}\geq 0$, 
\begin{equation}
\varepsilon _{K}\equiv \mathcal{E}_{K}\text{ }{-}\text{ }\mathcal{E}{_{0}}=%
\frac{2}{\pi }\hbar v_{F}K+\left[ 1-(2-[4/\pi ]^{2})\frac{E_{F}}{B_{2}}%
\right] \frac{\hbar ^{2}K^{2}}{2(2m)}+O(K^{3}),  \label{7}
\end{equation}%
where a nonnegative \textit{CP excitation energy }$\varepsilon _{K}$ has
been defined, and the Fermi velocity $v_{F}$ comes from $E_{F}/k_{F}=\hbar
v_{F}/2$. The leading term in (\ref{7}) is linear in $K$, followed by a
quadratic term. It is clear that the leading term in (\ref{7}) is quadratic,
namely 
\begin{equation}
\varepsilon _{K}=\hbar ^{2}K^{2}/2(2m)+O(K^{3}),  \label{8}
\end{equation}%
\textit{provided }$v_{F}$ and hence $E_{F}$ vanish, i.e., there is no Fermi
sea.\ This is just the familiar nonrelativistic kinetic energy in vacuum of
the composite (so-called ``local'') pair of mass $2m$ and CMM $K$. The same
result (\ref{8}) is also found to hold in 3D but not analytically as here in
2D. Figure 1 shows exact numerical results of a dimensionless CP excitation
energy $\varepsilon {_{K}/(-\mathcal{E}}_{0}{)}$ (in figure, ${\Delta }_{0}$
means present ${-}$ $\mathcal{E}{_{0}}>0$)\ as function of $K/k_{F}$ for
different couplings $B_{2}$. Note that ordinary CPs \textit{break up}
whenever $\mathcal{E}_{K\text{ }}$ turns from negative to positive, i.e.,
when $\mathcal{E}_{K\text{ }}$ vanishes, or by (\ref{7}) when $\varepsilon {%
_{K}/(-}\mathcal{E}_{0})\equiv \varepsilon {_{K}/\Delta }_{0}$ ${=1}$. These
points are marked in the figure by dots. In addition to the exact results
(full curves) also shown are some results for the linear approximation
[first term on the right-hand side of (\ref{7}), dot-dashed lines (virtually
coinciding with the exact curve for all $B_{2}/E_{F}\lesssim 0.1$)], as well
as for the quadratic approximation (dashed parabolas) as given by the
leading term in (\ref{8}) for stronger couplings. For weak enough coupling
the exact dispersion relation is virtually linear---\textit{in spite of the
divergence} of the quadratic term in (\ref{7}) as $B_{2}/E_{F}\rightarrow 0$%
. As $E_{F}$ decreases the quadratic dispersion relation (\ref{8}) very
slowly begins to dominate. A result unique to 2D (and associated with the
fact that in 2D the fermionic density of states is \textit{independent }of
energy) is that ${-}$ $\mathcal{E}{_{0}\equiv \Delta }_{0}=B_{2}$.

In 3D \cite{PhysicaC} instead of (\ref{6}) similar procedures for
two-component fermions give 
\begin{equation}
\sum_{\mathbf{k}}\frac{1}{\hbar ^{2}k^{2}/m}\text{ }-\sum_{\mathbf{k},(|%
\mathbf{k}\pm \mathbf{K}/2|>k_{F})}\frac{1}{\hbar ^{2}k^{2}/m-\mathcal{E}%
_{K}-2E_{F}+\hbar ^{2}K^{2}/4m}=\frac{mL^{3}}{4\pi \hbar ^{2}}\frac{1}{a}
\label{9}
\end{equation}%
where $a$ is the $s$-wave scattering length associated with the regularized $%
\delta $-well, which corresponds with weak to strong coupling according as $%
-\infty <1/k_{F}a<+\infty $. One finds for weak coupling ($k_{F}a\rightarrow
0^{-}$, e.g., prior to the well-known first-bound-state singularity as the
depth of a 3D potential well is increased) that 
\begin{equation}
\mathcal{E}_{0}/E_{F}\rightarrow -(8/e^{2})\exp (-\pi /k_{F}|a|),  \label{10}
\end{equation}%
a result first reported by Van Hove \cite{vanhove}. For strong coupling ($%
k_{F}a\rightarrow 0^{+}$, beyond the single-bound-state resonance) one gets%
\begin{equation}
\mathcal{E}_{0}/E_{F}\rightarrow -2/(k_{F}a)^{2}.  \label{11}
\end{equation}%
Numerical results in 3D very similar to those in Fig. 1 for 2D are obtained.
Namely, for weak coupling the CP dispersion curves are very nearly linear
while for smaller density they very slowly tend to the quadratic. The limit
given by (\ref{7}) in 2D was found too complicated in 3D to be evaluated
analytically, except for weak coupling. Repeating the 2D analysis without
attempting to explicitly determine the coefficient of the quadratic term one
gets, for the $\delta $-well interfermion interaction in weak coupling, 
\begin{equation}
\varepsilon {_{K}}\equiv \mathcal{E}_{K}\text{ }{-}\text{ }\mathcal{E}{_{0}}=%
\frac{1}{2}\hbar v_{F}K+O(K^{2}).  \label{12}
\end{equation}%
This is the same result cited without proof by Schrieffer in 1964 (Ref. \cite%
{Schrieffer}, p. 33)\ for the BCS model interaction, to which we now turn.

\subsection{BCS model interaction}

The BCS model interaction is of the simple form%
\begin{equation}
V_{\mathbf{kk^{\prime }}}=\left\{ 
\begin{array}{cl}
-V\, & \text{if \ }k_{F}<|\mathbf{k}\pm \frac{1}{2}\mathbf{K}|,\;|\mathbf{k}%
^{\prime }\pm \frac{1}{2}\mathbf{K}|<\sqrt{k_{F}^{2}+k_{D}^{2}} \\ 
0 & \text{otherwise,}%
\end{array}%
\right.   \label{13}
\end{equation}%
where $V_{\mathbf{kk^{\prime }}}$\ is defined in (\ref{4a}). Here $V>0$, and 
$\hbar \omega _{D}\equiv \hbar ^{2}k_{D}^{2}/2m$ is the maximum energy of a
vibrating-ionic-lattice phonon. This means that two fermions interact with a
constant attraction $-V$ when the tip of their relative-momentum wavevector $%
\mathbf{k}$ points anywhere inside the overlap volume in $k$-space of the
two spherical shells in Figure 2. Inserting (\ref{13}) into (\ref{4}) and
converting sums over $\mathbf{k}$\ into energy integrals by introducing the
electronic density of states (DOS) $g(\epsilon )$ gives 
\begin{equation}
1=V{\sum_{\mathbf{k}}}^{\prime }[2\epsilon _{k}-2E_{F}-\mathcal{E}_{K}+\hbar
^{2}K^{2}/4m]^{-1}=V\int_{E_{F}}^{E_{F}+\hbar \omega _{D}}{\frac{g(\epsilon
)d\epsilon }{2\epsilon -2E_{F}-\mathcal{E}_{K}+\hbar ^{2}K^{2}/4m}}.
\label{eq:cooper}
\end{equation}%
>From this one immediately obtains the familiar result for $K=0$,%
\begin{equation}
\mathcal{E}_{0}=-{\frac{2\hbar \omega _{D}}{e^{2/\lambda }-1}}\;\;%
\mathrel{\mathop{\longrightarrow}\limits_{\lambda \rightarrow 0}}\;\;-2\hbar
\omega _{D}e^{-2/\lambda }.  \label{eqn:Delta0}
\end{equation}%
Here $\lambda \equiv g(E_{F})V$ is a dimensionless coupling constant and $%
g(E_{F})$ the DOS for each spin evaluated at the Fermi energy. The equality
in (\ref{eqn:Delta0}) is \textit{exact} in 2D for all coupling---as well as
in 1D or 3D provided only that $\hbar \omega _{D}$ $\ll E_{F}$ so that $%
g(\epsilon )\simeq g(E_{F})$, a constant that can be taken outside the
integral in (\ref{eq:cooper}).

For a 2D system (\ref{eq:cooper}) gives \cite{PC98}\ for weak coupling%
\begin{equation}
\mathcal{E}_{K}\ \ \mathrel{\mathop{\longrightarrow}\limits_{K \rightarrow
0}}\ \ \mathcal{E}_{0}+({2/}\pi )\hbar v_{F}K+O(K^{2}).  \label{DeltaK0}
\end{equation}%
The exact dispersion relation obtained numerically from (\ref{eq:cooper})\
for $\lambda =%
{\frac12}%
$ and $\hbar \omega _{D}/E_{F}=10^{-2}$ shows that the linear approximation (%
\ref{DeltaK0}) is very good for moderately small $\lambda $ and $\hbar
\omega _{D}/E_{F}$, over the entire range of $K$ values for which $\mathcal{E%
}_{K}\leq 0$. Note that the linear term carries the \textit{same }%
coefficient as (\ref{7}) for a \textit{different }interfermion interaction.
Pair breakup, specifically $\mathcal{E}_{K}>0$ for these values of $\lambda $
and $\hbar \omega _{D}/E_{F}$, occurs at a relatively small value of $K$,
about four orders of magnitude smaller than the maximum value $2\sqrt{%
k_{F}^{2}+k_{D}^{2}}$ allowed by the interaction (\ref{13}).

In 3D, assuming $\hbar \omega _{D}/E_{F}<<1$ so that the 3D DOS $%
g(\varepsilon )=(L^{3}/\pi ^{2}\hbar ^{3})\sqrt{m^{3}\varepsilon /2}$ can be
replaced by $g(E_{F})$ and then taken outside the integral sign, the result
cited in Ref. \cite{Schrieffer}, p. 33, (see also Ref. \cite{FW}, p. 336,
Prob. 10.4 but note here an erroneous coeffcient) follows, namely%
\begin{equation}
\mathcal{E}_{K}\;\;\mathrel{\mathop{\longrightarrow}\limits_{K \rightarrow
0}}\;\;\mathcal{E}_{0}+{\frac{1}{2}}\hbar v_{F}K+O(K^{2}).  \label{DeltaK}
\end{equation}%
Exact numerical results in 3D are qualitatively similar \cite{PC98}\ to
those in 2D as regards goodness of the linear approximation for weak
coupling.

\section{Generalized Cooper pairing}

The ``ordinary'' CP problem just summarized for two distinct interfermion
interactions (the $\delta $-well and the BCS model interaction) neglects the
effect of two-hole (2h) CPs treated on an equal footing with two-particle
(2p) CPs---as Green's functions \cite{FW}\ can naturally guarantee. On the
other hand, the BCS condensate consists \cite{BF7a} of \textit{equal numbers
of 2p and 2h Cooper ``correlations}.'' This was already evident, though
scarcely emphasized, from the perfect symmetry\ about $\mu $, the electron
chemical potential, of the well-known Bogoliubov \cite{Bog} $v^{2}(\epsilon )
$\ and $u^{2}(\epsilon )$ coefficients [see just below (\ref{BCSgreen})
later on], where $\epsilon $ is the electron energy.\ the prime motivation
here rests on the recently established remarkable fact \cite{BF7a}\ that a
BCS condensate is precisely a BE condensate with equal numbers of 2p and 2h
zero-CMM CPs, in the limit of weak coupling. Further motivation comes from
the unique but unexplained role played by \textit{hole }charge carriers in
the normal state of superconductors \cite{Chapnik}\ in general (see also
Ref. \cite{Hirsch}). Final motivation stems from the ability of the
``complete (in that both 2h- and 2p-CPs are allowed in varying proportions)
BF model'' of Refs. \cite{BF7a}-\cite{CMT02} to ``unify'' both BCS and BEC
theories as special cases, and to predict substantially higher $T_{c}$'s
than BCS theory without abandoning electron-phonon dynamics. The latter is
important as compelling evidence for a significant, if not sole, presence of
it in high-$T_{c}$ cuprate superconductors from angle-resolved photoemission
spectroscopy data has recently been reported \cite{Shen}.

In this section we sketch how the Bethe-Salpeter (BS) many-body equation (in
the ladder approximation)\ treating both 2p and 2h pairs on an equal footing
shows that, while the ordinary CP problem [based on an ideal Fermi gas (IFG)
ground state (the usual ``Fermi sea'')] does \textit{not} possess stable
energy solutions, but that it does so when the IFG ground state is replaced
by the BCS one. This is equivalent to starting from an unperturbed
Hamiltonian that is the BCS ground state instead of the pure-kinetic-energy
operator corresponding to the IFG. We discuss how: i) CPs based not on the
IFG-sea but on the BCS ground state survive in a \textit{nontrivial}
solution as ``generalized'' or ``moving''\ CPs which are \textit{positive}
energy resonances with an imaginary energy term leading to finite-lifetime
effects; ii) as in the ``ordinary'' CP problem of the previous section,
their dispersion relation in leading order in the total (or center-of-mass)
momentum (CMM) $\hbar \mathbf{K\equiv \hbar (k}_{1}+\mathbf{k}_{2})$ is also 
\textit{linear }rather than the quadratic $\hbar ^{2}K^{2}/2(2m)$ of a
composite boson (e.g., a deuteron) of mass $2m$\ moving not in the Fermi sea
but in vacuum; and iii) this latter ``moving CP'' solution, though often
confused \cite{Randeria96}\ with it, is physically \textit{distinct }from
another more common \textit{trivial} solution sometimes called, even though
Bogoliubov \cite{Bog58}\ was the first to derive it, the
Anderson-Bogoliubov-Higgs (ABH) \cite{ABH}, (\cite{bts} p. 44), \cite{Higgs}-%
\cite{Traven2} collective excitation. The ABH mode is also linear in leading
order and goes over into the IFG ordinary sound mode in zero coupling. All
this occurs in both 2D \cite{ANFdeLl} as well as in the 3D study outlined
earlier in Ref. \cite{Honolulu}. We focus here on 2D because of its interest %
\cite{Varma}-\cite{Brandow}\ for quasi-2D high-$T_{c}$ cuprate
superconductors. In general, the results will be crucial for BEC scenarios
employing BF models of superconductivity, not only \textit{in} \textit{%
exactly 2D} as with the Berezinskii-Kosterlitz-Thouless \cite{BKT}\cite{KT}
transition, but also down to ($1+\epsilon $)D which characterize the
quasi-1D organo-metallic (Bechgaard salt) superconductors \cite%
{organometallics}-\cite{jerome2}. These results also apply, albeit with a
different interaction, to neutral-atom superfluidity as in liquid $^{3}$He\ %
\cite{He3} and very probably also in ultracold trapped alkali Fermi gases
such as $^{40}$K \cite{Holland} and $^{6}$Li \cite{Li6} atoms.

In dealing with the many-electron system we again assume the BCS model
interaction (\ref{4a}) in 2D with double Fourier transform 
\begin{eqnarray}
\nu (k_{1},k_{1}^{\prime }) &=&-(k_{F}^{2}/k_{1}k_{1}^{\prime })V\text{ \ \
if}\;\ k_{F}-k_{D}<k_{1},k_{1}^{\prime }<k_{F}+k_{D},  \notag \\
\  &=&0\ \ \ \text{otherwise.}  \label{int}
\end{eqnarray}%
Here $V>0$, $\hbar k_{F}\equiv mv_{F}$ the Fermi momentum, $m$ the effective
electron mass, $v_{F}$ the Fermi velocity, and $k_{D}\equiv \omega _{D}/v_{F}
$  with $\omega _{D}$ the Debye frequency; note difference with previous
definition just below (\ref{13}). The usual physical constraint $\hbar
\omega _{D}\ll E_{F}$ then implies that $k_{D}/k_{F}\equiv \hbar \omega
_{D}/2E_{F}\ll 1$. Assuming perfect ph symmetry about the Fermi surface, we
set%
\begin{equation}
\epsilon _{k}\simeq E_{F}+\hbar v_{F}(k-k_{F})  \label{14}
\end{equation}%
as it simplifies all calculations when very near the Fermi surface.

The bound-state BS wavefunction equation \cite{Honolulu} in the ladder
approximation\ with both particles and holes for the original IFG-based CP
problem\ is%
\begin{eqnarray}
\Psi (\mathbf{k,}E)=&-\left( \frac{i}{\hbar }\right) ^{2}G_{0}\left( \mathbf{K%
}/2+\mathbf{k},\mathcal{E}_{K}/2+E\right) G_{0}\left( \mathbf{K}/2-\mathbf{k}%
,\mathcal{E}_{K}/2-E\right) \times   \notag \\
&\hspace{-1.0cm}\times \frac{1}{2\pi i}\int\limits_{-\infty }^{+\infty }dE^{^{\prime }}%
\frac{1}{L^{d}}\sum_{\mathbf{k}^{\prime }}v(\left| \mathbf{k-k}^{\prime
}\right| )\Psi (\mathbf{k}^{\prime },E^{\prime }).  \label{BSE}
\end{eqnarray}
Here $L^{d}$ is the ``volume''\ of the $d$-dimensional system; $\mathbf{%
K\equiv k}_{1}+\mathbf{k}_{2}$ is the CMM and $\mathbf{k\equiv 
{\frac12}%
(k}_{1}-\mathbf{k}_{2}\mathbf{)}$ the relative wavevectors of the 2e bound
state whose wavefunction is $\Psi (\mathbf{k,}E)$; $\mathcal{E}_{K}$ $\equiv
E_{1}+E_{2}$ is the energy of this bound state while $E\equiv E_{1}-E_{2}$,
and $G_{0}\left( \mathbf{K}/2+\mathbf{k},\mathcal{E}/2+E\right) $ is the
bare one-fermion Green's function given by%
\begin{equation}
G_{0}(\mathbf{k}_{1},E_{1})=\frac{\hbar }{i}\left\{ \frac{\theta
(k_{1}-k_{F})}{-E_{1}+\epsilon _{\mathbf{k}_{1}}-E_{F}-i\varepsilon }+\frac{%
\theta (k_{F}-k_{1})}{-E_{1}+\epsilon _{\mathbf{k}_{1}}-E_{F}+i\varepsilon }%
\right\}   \label{IFGgreen}
\end{equation}%
where $\epsilon _{\mathbf{k}_{1}}\equiv $ $\hbar ^{2}k_{1}^{2}/2m$ and $%
\theta (x)=1$ for $x>0$ and $=0$ for $x<0$, so that the first term refers to 
\textit{electrons} and the second to \textit{holes}. Figure 3 shows all
Feynman diagrams for the 2p, 2h and ph wavefunctions $\psi _{+}$, $\psi _{-}$
and $\psi _{0}$, respectively, that emerge in the general
(BCS-ground-state-based) problem to be discussed later. For the present
IFG-based case, diagrams in shaded rectangles do \textit{not} contribute as
they involve factors of $\theta (k_{1}-k_{F})\theta (k_{F}-k_{1})\equiv 0$.
Since the energy dependence of $\Psi (\mathbf{k},E)$ in (\ref{BSE}) is only
through the Green's functions, the ensuing energy integrals (\ref{BSE})\ can
be evaluated directly in the complex $E^{\prime }$-plane and yield, for
interaction (\ref{int}), an equation for the wavefunction $\psi _{\mathbf{k}}
$ in momentum space for CPs with \textit{zero}\ CMM $\mathbf{K\equiv k}_{1}+%
\mathbf{k}_{2}=0$ that is 
\begin{equation}
(2\xi _{k}-\mathcal{E}_{0})\psi _{\mathbf{k}}=V\sum_{\mathbf{k}^{\prime
}}{}^{^{\prime }}\psi _{\mathbf{k}^{\prime }}-V\sum_{\mathbf{k}^{\prime
}}{}^{^{\prime \prime }}\psi _{\mathbf{k}^{\prime }}.  \label{CP}
\end{equation}%
Here $\xi _{k}\equiv \hbar ^{2}k^{2}/2m-E_{F}$ while $\mathcal{E}_{0}$ is
the $\mathbf{K}=0$\ eigenvalue energy, and $\mathbf{k\equiv 
{\frac12}%
({k}_{1}-{k}_{2})=k}_{1}$. The single prime over the first (2p-CP) summation
term denotes the restriction $0<\xi _{k^{\prime }}<\hbar \omega _{D}$ while
the double prime in the last (2h-CP) term means $-\hbar \omega _{D}<\xi
_{k^{\prime }}<0$. Without this latter term we have Cooper's Schr\"{o}%
dinger-like equation \cite{Coo}\ for 2p-CPs whose implicit solution is
clearly $\psi _{\mathbf{k}}=(2\xi _{k}-\mathcal{E}_{0})^{-1}V\sum_{\mathbf{k}%
^{\prime }}^{^{\prime }}\psi _{\mathbf{k}^{\prime }}.$ Since the summation
term is constant, performing that summation on both sides allows canceling
the $\psi _{\mathbf{k}}$-dependent terms, leaving the eigenvalue equation $%
\sum_{\mathbf{k}}^{^{\prime }}(2\xi _{k}-\mathcal{E}_{0})^{-1}=1/V$. This is
one equation in one unknown $\mathcal{E}_{0}$; transforming the sum to an
integral over energies gives the familiar solution (\ref{eqn:Delta0}) exact
in 2D, and to a very good approximation otherwise if $\hbar \omega _{D}\ll
E_{F}$, where $\lambda \equiv VN(E_{F})$ with $N(E_{F})$ the electronic DOS
for one spin. This corresponds to a negative-energy, stationary-state bound
pair. For $K\geqslant 0$ the 2p-CP eigenvalue equation becomes 
\begin{equation}
\sum_{\mathbf{k}}{}^{^{\prime }}(2\xi _{k}-\mathcal{E}_{K}+\hbar
^{2}K^{2}/4m)^{-1}=1/V.  \label{CPKeqn}
\end{equation}%
Note that a 2p CP state of energy $\mathcal{E}_{K}$\ is characterized only
by a definite $K$ but \textit{not }definite $\mathbf{k}$, in contrast to a
``BCS pair'' defined [Ref. \cite{bcs}, Eqs. (2.11) to (2.13)]\ with fixed\ $%
\mathbf{K}$ and $\mathbf{k}$ (or equivalently definite $\mathbf{k}_{1}$\ and 
$\mathbf{k}_{2}$); see Appendix A. Without the first summation term in (\ref%
{CP})\ the same expression (\ref{eqn:Delta0})\ for $\mathcal{E}_{0}$\ for
2p-CPs follows for 2h-CPs, apart from an overall sign change.

The \textit{complete }equation (\ref{CP}) \textit{cannot }be derived from an
ordinary (non-BS) Schr\"{o}dinger-like equation in spite of its simple
appearance. To solve it for the unknown energy $\mathcal{E}_{0}$, let the
rhs of (\ref{CP}) be defined as $A-B$, with $A$ relating to the 2p term and $%
B$ to the 2h one. Solving for the unknown $\psi _{\mathbf{k}}$ gives%
\begin{equation}
\psi _{\mathbf{k}}=(A-B)/(2\xi _{k}-\mathcal{E}_{0})\text{ \ or equivalently
\ }\psi (\xi )=(A-B)/(2\xi -\mathcal{E}_{0})  \label{15}
\end{equation}%
whence 
\begin{eqnarray}
A &\equiv &\lambda \int_{0}^{\hbar \omega _{D}}d\xi \psi (\xi )=%
{\frac12}%
(A-B)\lambda \int_{-\mathcal{E}_{0}}^{2\hbar \omega _{D}-\mathcal{E}%
_{0}}dz/z\equiv (A-B)x  \label{16} \\
B &\equiv &\lambda \int_{-\hbar \omega _{D}}^{0}d\xi \psi (\xi )=%
{\frac12}%
(A-B)\lambda \int_{-2\hbar \omega _{D}-\mathcal{E}_{0}}^{-\mathcal{E}%
_{0}}dz/z\equiv (A-B)y.  \label{17}
\end{eqnarray}%
The integrals are readily evaluated giving $x\equiv 
{\frac12}%
\lambda \ln (1-2\hbar \omega _{D}/\mathcal{E}_{0})$ and $y\equiv -%
{\frac12}%
\lambda \ln (1+2\hbar \omega _{D}/\mathcal{E}_{0})$. As $A$ and $B$ still
contain the unknown $\psi (\xi )$ let us eliminate them. Note that (\ref{16}%
) and (\ref{17}) are equivalent to \textit{two} equations in two unknowns $A$
and $B$, namely 
\begin{eqnarray*}
(1-x)A+xB &=&0 \\
-yA+(1+y)B &=&0.
\end{eqnarray*}%
This leads immediately to the equation $1-x+y=0,$ which on inserting the
definitions of $x$ and $y$ becomes 
\begin{equation*}
1=%
{\frac12}%
\lambda \ln [1-(2\hbar \omega _{D}/\mathcal{E}_{0})^{2}],
\end{equation*}%
which finally yields 
\begin{equation}
\mathcal{E}_{0}=\pm i2\hbar \omega _{D}/\sqrt{e^{2/\lambda }-1}.  \label{18}
\end{equation}%
As the CP energy is pure-imaginary there is an obvious instability of the CP
problem when both particle- and hole-pairs are included. This was reported
in Refs. \cite{bts} p. 44 and \cite{AGD}, who did not, however, stress the
pure 2p and 2h special cases just discussed. Clearly then, the original CP
picture \textit{is meaningless if particle- and hole-pairs are treated on an
equal footing}, as consistency demands.

\section{BCS-based BS treatment of Cooper pairing}

However, a BS treatment not about the IFG sea but about the BCS ground state 
\textit{vindicates the CP concept }as a nontrivial solution. This is
equivalent to starting not from the IFG unperturbed Hamiltonian but from the
BCS one. Its physical justification lies in recovering three expected items:
ABH sound mode, the BCS $T=0$ gap equation\ and finite-lifetime effects of
the ``moving CPs.'' In either 2D \cite{ANFdeLl} or 3D \cite{Honolulu}\ the
BS equation yields a $4\times 4$ determinant which reduces to a $3\times 3$
and a $1\times 1$ determinant representing, respectively, \textit{two} 
\textit{distinct} \textit{solutions}: a) the trivial ABH sound solution and
b) a highly \textit{nontrivial} moving CP solution, respectively. In either
case the BS formalism gives a set of three coupled equations, one for each
(2p, 2h and ph) channel\ wavefunction\ for any spin-independent interaction
such as (\ref{int}). However, the ph channel decouples, leaving only two
coupled wavefunction\ equations for the ABH solution in 2D which we consider
first. We note that in Ref. \cite{Ran} the hh channel was ignored, leading
to a $3\times 3$ determinant\ from which only the trivial ABH solution
emerges, but the nontrivial moving CP one was missed entirely. Thus, the IFG
Green function (\ref{IFGgreen}) is replaced by the BCS one 
\begin{equation}
\text{\textbf{G}}_{0}(\mathbf{k}_{1},E_{1})=\frac{\hbar }{i}\left\{ \frac{%
v_{k_{1}}^{2}}{-E_{1}+E_{k_{1}}-i\varepsilon }+\frac{u_{k_{1}}^{2}}{%
-E_{1}+E_{k_{1}}+i\varepsilon }\right\}   \label{BCSgreen}
\end{equation}%
where $E_{\mathbf{k}}\equiv \sqrt{\xi _{k}{}^{2}+\Delta ^{2}}$ with $\Delta $
the $T=0$ fermionic gap, $v_{k}^{2}\equiv 
{\frac12}%
(1-\xi _{k}/E_{\mathbf{k}})$ and $u_{k}^{2}\equiv 1-v_{k}^{2}$ are the
Bogoliubov functions \cite{Bog47}.\ As $\Delta \rightarrow 0$\ these three
quantities become $|\xi _{k}|$, $\theta (k_{1}-k_{F})$ and $\theta
(k_{F}-k_{1})$, respectively, so that (\ref{BCSgreen}) reduces to (\ref%
{IFGgreen}),\ as expected. Substituting $G_{0}(\mathbf{k}_{1},E_{1})$ by 
\textbf{G}$_{0}(\mathbf{k}_{1},E_{1})$\ corresponds to rewriting the total
Hamiltonian so that the pure-kinetic-energy unperturbed Hamiltonian is
replaced by the BCS one. The remaining Hamiltonian terms are then assumed
suitable to a perturbation treatment. We focus in this section only on 2D.

\subsection{ABH sound (trivial) solution\protect\bigskip}

The equations involved are too lengthy even in 2D\ and will be derived in
detail elsewhere, but for the trivial ABH sound solution the aforementioned $%
3\times 3$ determinant boils down to the single expression 
\begin{gather}
\frac{1}{2\pi }\lambda \hbar
v_{F}\int_{k_{F}-k_{D}}^{k_{F}+k_{D}}dk\int_{0}^{2\pi }d\varphi \{u_{\mathbf{%
K}/2+\mathbf{k}}{\Huge \,}u_{\mathbf{K}/2-\mathbf{k}}\text{ }+\text{ }v_{%
\mathbf{K}/2+\mathbf{k}}{\Huge \,}v_{\mathbf{K}/2-\mathbf{k}}\}\times  
\notag \\
\times {\Huge [}\frac{v_{\mathbf{K}/2+\mathbf{k}}v_{\mathbf{K}/2-\mathbf{k}}%
}{\mathcal{E}_{K}+E_{\mathbf{K}/2+\mathbf{k}}\text{ }+\text{ }E_{\mathbf{K}%
/2-\mathbf{k}}}+\,\frac{u_{\mathbf{K}/2+\mathbf{k}}\,u_{\mathbf{K}/2-\mathbf{%
k}}}{-\mathcal{E}_{K}+E_{\mathbf{K}/2+\mathbf{k}}+E_{\mathbf{K}/2-\mathbf{k}}%
}{\Huge ]}=1  \label{19}
\end{gather}%
where $\varphi $ is the angle between $\mathbf{K}$ and $\mathbf{k}$. Here $%
k_{D}\equiv \omega _{D}/v_{F}$; note difference with definition just below (%
\ref{13}). As before $\lambda \equiv VN(E_{F})$ with $N(E_{F})\equiv m/2\pi
\hbar ^{2}$ the constant 2D electronic DOS and $V>0$ is the interaction
strength defined in (\ref{int}). Angle-resolved photoemission spectral
studies of $BiSrCaCuO$ have shown evidence \cite{Matsui03} in this cuprate
for the Bogoliubov functions\ $u_{k}^{2}$ and $v_{k}^{2}$, both above and
below the Fermi energy.

The \textit{ABH collective excitation mode }energy $\mathcal{E}_{K}$\ must
then be extracted from this equation. For $\mathbf{K}=0$ it is just $%
\mathcal{E}_{0}=0$ (Ref. \cite{bts} p. 39). Then (\ref{19}) rewritten as an
integral over $\xi \equiv \hbar ^{2}k^{2}/2m-E_{F}$ reduces to%
\begin{equation}
\int_{0}^{\hbar \omega _{D}}d\xi /\sqrt{\xi ^{2}+\Delta ^{2}}=1/\lambda ,
\label{20}
\end{equation}%
or the familiar BCS $T=0$ gap equation for interaction (\ref{int}). The
integral is exact and gives%
\begin{equation}
\Delta =\ \hbar \omega _{D}/\sinh (1/\lambda ).  \label{21}
\end{equation}%
Returning to the ABH energy $\mathcal{E}_{K}$ equation (\ref{19})\ and
Taylor-expanding $\mathcal{E}_{K}$ about $K=0$ and small $\lambda $ leaves 
\begin{equation}
\mathcal{E}_{K}=\frac{\hbar v_{F}}{\sqrt{2}}K+O(K^{2})+o(\lambda ),
\label{22}
\end{equation}%
where $o(\lambda )$ denote interfermion interaction correction terms that
vanish as $\lambda \rightarrow 0$.\ Note that the leading term is just the
ordinary sound mode in an IFG whose sound speed $c$ $=$ $v_{F}/\sqrt{d}$ in $%
d$ dimensions.

The latter also follows elementarily on solving for $c$ in the familiar
thermodynamic relation $dP/dn=mc^{2}$ involving the zero-temperature IFG
pressure 
\begin{eqnarray}
P &=&n^{2}[d(E/N)/dn]=2nE_{F}/(d+2)  \notag \\
&=&2C_{d}n^{2/d\text{ }+1}/(d+2)  \label{23}
\end{eqnarray}%
where the constant $C_{d}$ will drop out. Here the IFG ground-state
(internal) energy per fermion $E/N=dE_{F}/(d+2)=C_{d}n^{2/d}$ was used along
with $E_{F}\equiv \hbar ^{2}k_{F}^{2}/2m$ and%
\begin{equation}
n\equiv N/L^{d}=k_{F}^{d}/d2^{d-2}\pi ^{d/2}\Gamma (d/2)  \label{24}
\end{equation}%
for the fermion-number density $n$. The derivative of (\ref{23})\ with
respect to $n$ finally gives $c$ $=$ $\hbar k_{F}/m\sqrt{d}\ \equiv v_{F}/%
\sqrt{d}$, which in 2D is just the leading term in (\ref{22}).

\subsection{Moving CP (nontrivial) solution}

The nontrivial \textit{moving CP}\ solution of the BCS-ground-state-based BS
treatment, which is \textit{entirely new}, comes from the remaining $1\times
1$\ determinant. It leads to the pair energy $\mathcal{E}_{K}$ which in 2D
is contained in the equation 
\begin{gather}
\frac{1}{2\pi }\lambda \hbar
v_{F}\int_{k_{F}-k_{D}}^{k_{F}+k_{D}}dk\int_{0}^{2\pi }d\varphi u_{\mathbf{K}%
/2+\mathbf{k}}v_{\mathbf{K}/2-\mathbf{k}}\times   \notag \\
\times \{u_{\mathbf{K}/2-\mathbf{k}}v_{\mathbf{K}/2+\mathbf{k}}-u_{\mathbf{K}%
/2+\mathbf{k}}v_{\mathbf{K}/2-\mathbf{k}}\}\frac{E_{\mathbf{K}/2+\mathbf{k}%
}+E_{\mathbf{K}/2-\mathbf{k}}}{-\mathcal{E}_{K}^{2}+(E_{\mathbf{K}/2+\mathbf{%
k}}+E_{\mathbf{K}/2-\mathbf{k}})^{2}}=1.  \label{mCP}
\end{gather}%
In addition to the pp and hh wavefunctions (depicted diagrammatically in
Ref. \cite{Honolulu} Fig. 2), diagrams associated with the ph channel give
zero contribution at $T=0$. A third equation for the ph wavefunction
describes the ph bound state but turns out to depend only on the pp and hh
wavefunctions. Taylor-expanding $\mathcal{E}_{K}$ in (\ref{mCP}) in powers
of $K$\ around $K=0$, and introducing a possible damping factor by adding an
imaginary term $-i\Gamma _{K}$ in the denominator, yields to order $K^{2}$%
\begin{align}
\pm \mathcal{E}_{K}& \simeq 2\Delta +\frac{\lambda }{2\pi }\hbar v_{F}K+%
\frac{1}{9}\frac{\hbar v_{F}}{k_{D}}e^{1/\lambda }K^{2}  \notag \\
& -i\left[ \frac{\lambda }{\pi }\hbar v_{F}K+\frac{1}{12}\frac{\hbar v_{F}}{%
k_{D}}e^{1/\lambda }K^{2}\right] +O(K^{3})  \label{linquadmCP}
\end{align}%
where the upper and lower signs refer to 2p- and 2h-CPs, respectively. A
linear dispersion in leading order again appears, but now associated with
the bosonic moving CP. Hence the \textit{positive}-energy 2p-CP resonance
has a width $\Gamma _{K}$ and a lifetime%
\begin{equation}
\tau _{K}\equiv \hbar /2\Gamma _{K}=\hbar /2\left[ (\lambda /\pi )\hbar
v_{F}K+(\hbar v_{F}/12k_{D})e^{1/\lambda }K^{2}\right] .  \label{25}
\end{equation}%
This diverges only at $K=0$, falling to zero as $K$ increases. Thus,
``faster'' moving CPs are shorter-lived and eventually break up, while
``non-moving'' ones are infinite-lifetime stationary states. The linear term 
$(\lambda /2\pi )\hbar v_{F}K$ in (\ref{linquadmCP}) contrasts sharply with
the \textit{coupling-independent} leading-term in (\ref{DeltaK0}) [or $1/2$
in 3D in (\ref{DeltaK}), Ref. \cite{Schrieffer} p. 33,\ instead of $2/\pi $]
that follows from the \textit{original }CP problem (\ref{CPKeqn})\
neglecting holes. This holds for either interaction (\ref{int}) \cite{PC98} 
\textit{or }for an attractive delta interfermion potential well in 2D \cite%
{PRB2000} or in 3D \cite{PhysicaC}. These $\delta $-wells are imagined
regularized \cite{GT} to possess a single bound level whose binding energy
(in 2D) or scattering length (in 3D) serve as the coupling parameter. Figure
4a shows\ the exact moving CP (mCP)\ energy (full curves) extracted from (%
\ref{mCP}), along with its leading linear-dispersion term (thin short-dashed
lines) and this plus the next (quadratic) term (long-dashed curves) from (%
\ref{linquadmCP}). The interaction parameter values used with (\ref{int})\
were $\hbar \omega _{D}/E_{F}=0.05$ (a typical value for cuprates) and the
two values $\lambda =%
{\frac14}%
$ and $%
{\frac12}%
$. Using (\ref{21}) in (\ref{linquadmCP}) gives%
\begin{equation}
\mathcal{E}_{0}/E_{F}\equiv 2\Delta /E_{F}=2\hbar \omega _{D}/E_{F}\sinh
(1/\lambda ),  \label{26}
\end{equation}%
having the values $\simeq 0.004$ and $0.028,$ respectively (marked as dots
on the figure ordinate). Remarkably enough, the linear approximation (thin
short-dashed lines in figure) is better over a wider range of $K/k_{F}$
values for weaker coupling (lower set of three curves) in spite of a larger
and larger (because of the factor $e^{1/\lambda }$) partial contribution
from the quadratic term in (\ref{linquadmCP}). This peculiarity also emerged
from the ordinary CP treatment of Sec. 3, Refs. \cite{PRB2000}\cite{PhysicaC}%
\cite{PC98}. It suggests the expansion in powers of $K$ to be an asymptotic
series that should be truncated after the linear term. For reference we also
plot the linear leading term $\hbar v_{F}K/\sqrt{2}$\ of the sound solution (%
\ref{22}). We note that the \textit{coupling-independent} leading-term \cite%
{PC98} $(2/\pi )\hbar v_{F}K$\ from the \textit{original }CP problem
neglecting holes, if graphed in Fig. 4, would almost coincide with the ABH
term $\hbar v_{F}K/\sqrt{2}$ and have a slope about $90\%$ smaller.

Empirical evidence for the \textit{linearly-dispersive} nature of Cooper
pairs in cuprates has been argued by Wilson \cite{Wilson03} to be suggested
by the scanning tunneling microscope conductance scattering data in BSCCO
obtained by Davis and coworkers \cite{Hoffman02}\cite{McElroy03}. More
suggestive direct evidence is shown in Figure 5 \cite{PRBcomm} with
experimental data (mostly from penetration-depth measurements)\ for two 3D
SCs \cite{Guimpel}\cite{Schawlow}, two quasi-2D cuprates \cite{Jacobs}-\cite%
{Bonn}, and a quasi-1D SC \cite{Tang}. The data\ are seen to agree quite
well, at least for $T\gtrsim 0.5T_{c}$, with the \textit{pure-phase} (only
2e- \textit{or} 2h-CP) BEC condensate fraction formula $1-(T/T_{c})^{d/s}$
for $d=3$, $2$ and $1$, respectively, \textit{provided one assumes }$s=1.$
For lower $T$'$s$, one can argue on the basis of Fig. 9 below that a \textit{%
mixed }BEC phase containing both 2e- and 2h-CPs becomes more stable (i.e.,
has lower Helmholtz free energy) so that the simple pure-phase $%
1-(T/T_{c})^{d/s}$ formula is no longer strictly valid.

As in Cooper's \cite{Coo}\ original equation (\ref{CPKeqn}), the BS moving
CPs are characterized by a definite $\mathbf{K}$ and \textit{not} also by
definite $\mathbf{k}$ as the pairs discussed by BCS \cite{bcs}. Hence, the
objection does not apply that CPs are not bosons because BCS pairs with
definite\ $\mathbf{K}$ and $\mathbf{k}$ (or equivalently definite $\mathbf{k}%
_{1}$\ and $\mathbf{k}_{2}$) have creation/annihilation operators that do 
\textit{not }obey the usual Bose commutation relations [Ref. \cite{bcs},
Eqs. (2.11) to (2.13)]. In fact, either (\ref{CPKeqn}) or (\ref{mCP}) shows
that a given ``ordinary'' or ``generalized'' CP state labeled by either $%
\mathbf{K}$ or $\mathcal{E}_{K}$\ can accommodate (in the thermodynamic
limit) an indefinitely many possible BCS pairs with different $\mathbf{k}$%
's; see Ref. \cite{BCSandCPs}\ or Appendix A. A recent electronic analog %
\cite{Samuelsson} of the Hanbury Brown-Twiss photon-effect experiment
suggest electron pairs in a SC to be definitely bosons.

To conclude this section, hole pairs treated on a par with electron pairs
play a vital role in determining the precise nature of CPs even at zero
temperature---only when based not on the usual IFG ``sea'' but on the BCS
ground state. Their treatment with a Bethe-Salpeter equation gives
purely-imaginary-energy CPs when based on the IFG, and positive-energy
resonant-state CPs with a finite lifetime for nonzero CMM when based on the
BCS ground state---instead of the more familiar negative-energy stationary
states of the original IFG-based CP problem that neglects holes, as sketched
just below (\ref{CP}). The BS ``moving-CP'' dispersion relation (\ref%
{linquadmCP}), on the other hand,\ resembles the plasmon dispersion curve in
3D. It is gapped by twice the BCS energy gap, followed by a \textit{linear}
leading term in the CMM expansion about $K=0$, instead of the quadratic for
the 3D plasmon curve. This linearity is distinct from the better-known one (%
\ref{22}) associated with the sound or ABH collective excitation mode whose
energy vanishes at $K=0$. Thus, BF models assuming this CP linearity for the
boson component, instead of the quadratic\ $\hbar ^{2}K^{2}/2(2m)$ assumed
in Refs. \cite{Blatt}, \cite{BF5}-\cite{PLA2}, \cite{Holland01}-\cite%
{Griffin02} among many others, can give BEC for all $d>1$, including\
exactly 2D. Such BF models can then in principle address not only quasi-2D
cuprate but also quasi-1D organo-metallic superconductors.

\section{The CBFM Hamiltonian}

The CBFM \cite{BF7a,PLA2} is described (in $d$ dimensions) by the
Hamiltonian $H=H_{0}+H_{int}$. The unperturbed Hamiltonian $H_{0}$\
corresponds to a ``normal'' state which is an \textit{ideal }(i.e.,
noninteracting) gas mixture of unpaired fermions and both types of CPs,
two-electron (2e) and two-hole (2h), and is given by%
\begin{equation}
H_{0}=\sum\limits_{\mathbf{k}_{1},s_{1}}\epsilon _{\mathbf{k}_{\mathbf{1}%
}}a_{\mathbf{k}_{1},s_{_{1}}}^{+}a_{\mathbf{k}_{1},s_{_{1}}}+\sum\limits_{%
\mathbf{K}}E_{+}(K)b_{\mathbf{K}}^{+}b_{\mathbf{K}}-\sum\limits_{\mathbf{K}%
}E_{-}(K)c_{\mathbf{K}}^{+}c_{\mathbf{K}}  \label{H0}
\end{equation}%
where as before $\mathbf{K\equiv k}_{\mathbf{1}}+\mathbf{k}_{\mathbf{2}}$ is
the CP CMM wavevector, $\mathbf{k}\equiv \frac{1}{2}\mathbf{(k}_{\mathbf{1}}-%
\mathbf{k}_{\mathbf{2}})$ its relative wavevector, while $\epsilon _{\mathbf{%
k}_{1}}\equiv \hbar ^{2}k_{1}^{2}/2m$ are the single-electron, and $E_{\pm
}(K)$\ the 2e-/2h-CP \textit{phenomenological, }energies.\ Here $a_{\mathbf{k%
}_{1},s_{1}}^{+}$ ($a_{\mathbf{k}_{1},s_{1}}$) are creation (annihilation)
operators for fermions and similarly $b_{\mathbf{K}}^{+}$ ($b_{\mathbf{K}}$)
and $c_{\mathbf{K}}^{+}$ ($c_{\mathbf{K}}$) for 2e- and 2h-CP bosons,
respectively. Two-hole CPs are considered \textit{distinct }and\textit{\
kinematically independent }from 2e-CPs since their Bose commutation
relations involve a relative sign change, in sharp contrast with electron or
hole fermions whose Fermi anticommutation relations do not.\ Our present
formulation is of course nonrelativistic because of relatively low
temperatures. 

At the opposite extreme of very high $T$'s (compared with the boson
rest-mass energy), however, one has a relativistic regime where pair
production takes place and BEC \textit{must }take $\overset{\_}{N}$
antibosons of charge, say, $-q$ into account along with the $N$ bosons of
charge $q$. In units such that $\hbar \equiv c\equiv k_{B}\equiv 1$ the
boson energy is $\varepsilon _{K}=(K^{2}+m_{B}^{2})^{1/2}$. Charge
conservation requires that not $N\equiv N_{0}(T)+\sum_{\mathbf{K\neq 0}%
}[\exp \beta (\varepsilon _{K}-\mu _{B})-1]^{-1}$ be constant but rather $N-$
$\overset{\_}{N\text{,}}$ where $\overset{\_}{N}$ is the same expression as $%
N$ but with $+\mu _{B}$ instead of $-\mu _{B}$. If $\rho \equiv q(N-\overset{%
\_}{N})/L^{3}\equiv qn$ is the net conserved charge density, it is shown in
Ref. \cite{HaberWeldon} that $T_{c}=(3|n|/m_{B})^{%
{\frac12}%
}$ and that the condensate fraction $n_{0}(T)/n=1-(T/T_{c})^{2}.$ This is 
\textit{qualitatively} different from the better-known results assuming only 
$N$ constant, which are the mass-independent $T_{c}=[\pi ^{2}n/\zeta
(3)]^{1/3}$, and $n_{0}(T)/n=1-(T/T_{c})^{3}.$ This analogy with the CBFM
exhibits the strikingly dramatic effect of including or not
antiparticles---or, in our nonrelativistic problem, of including or not hole
pairs.

The interaction Hamiltonian $H_{int}$ consists of four distinct BF
interaction vertices, see Fig. 6, each with two-fermion/one-boson creation
or annihilation operators, depicting how unpaired electrons (subindex +) [or
holes (subindex $-$)] combine to form the 2e- (and 2h-CPs) assumed in the $d$%
-dimensional system of size $L$, namely 
\begin{equation*}
H_{int}=L^{-d/2}\sum\limits_{\mathbf{k},\mathbf{K}}f_{+}(k)\{a_{\mathbf{k}+%
\frac{1}{2}\mathbf{K},\uparrow }^{+}a_{-\mathbf{k}+\frac{1}{2}\mathbf{K}%
,\downarrow }^{+}b_{\mathbf{K}}\text{ }+\text{ }a_{-\mathbf{k}+\frac{1}{2}%
\mathbf{K},\downarrow }a_{\mathbf{k}+\frac{1}{2}\mathbf{K},\uparrow }b_{%
\mathbf{K}}^{+}\}
\end{equation*}%
\begin{equation}
+L^{-d/2}\sum\limits_{\mathbf{k},\mathbf{K}}f_{-}(k)\{a_{\mathbf{k}+\frac{1}{%
2}\mathbf{K},\uparrow }^{+}a_{-\mathbf{k}+\frac{1}{2}\mathbf{K},\downarrow
}^{+}c_{\mathbf{K}}^{+}\text{ }+\text{ }a_{-\mathbf{k}+\frac{1}{2}\mathbf{K}%
,\downarrow }a_{\mathbf{k}+\frac{1}{2}\mathbf{K},\uparrow }c_{\mathbf{K}}\}.
\label{Hint}
\end{equation}%
Note that the \textit{fermion-pair interaction} $H_{int}$ is reminiscent%
\textit{\ }of\textit{\ }the Fr\"{o}hlich (or Dirac\ QED) interaction
Hamiltonian (Ref. \cite{FW}, p. 396 ff.) involving two fermion and one boson
operators, but with \textit{two}\textbf{\ }types of CPs instead of phonons
(or photons). But in contrast with Fr\"{o}hlich and Dirac there is no a
conservation law for the number of unpaired electrons, i.e.,$\
[H_{int},\sum\limits_{\mathbf{k}_{1},s_{1}}\varepsilon _{\mathbf{k}_{\mathbf{%
1}}}a_{\mathbf{k}_{1},s_{_{1}}}^{+}a_{\mathbf{k}_{1},s_{_{1}}}]\neq 0.$
(Note however that{\Huge {\large \ }}$[H_{int},\sum\limits_{\mathbf{k}%
_{1},s_{1}}\mathbf{k}_{1}a_{\mathbf{k}_{1},s_{_{1}}}^{+}a_{\mathbf{k}%
_{1},s_{_{1}}}]=0$ and $[H_{int},\sum\limits_{\mathbf{k}_{1},s_{1}}s_{1}a_{%
\mathbf{k}_{1},s_{_{1}}}^{+}a_{\mathbf{k}_{1},s_{_{1}}}]=0$.){\Huge \ }Just
as the Fr\"{o}hlich (or Dirac) interaction Hamiltonians are the most natural
ones to use in a many-electron/phonon (or photon) system, one can conjecture
the same of (\ref{Hint}) for the BF system under study. Indeed, this{\large %
\ }$H_{int}$ has \textit{formally} already been employed under various
guises by several authors \cite{BF3}-\cite{BF6},\cite{BF4}-\cite{BF8}, but 
\textit{without }hole pairs. More recently, a similar BF $H_{int}$ has been
employed \cite{Holland01}-\cite{Griffin02}\ to study quantum degenerate
Fermi gases consisting of neutral $^{40}K$\ atoms and their so-called
Feshbach ``resonance superfluidity.'' However, these authors assume
quadratically-dispersive CPs, besides also \textit{excluding 2h-CPs} and
cannot thus relate their formalism to BCS theory.

The energy form factors $f_{\pm }(k)$ in (\ref{Hint}) are\ essentially the
Fourier transforms of the 2e- and 2h-CP intrinsic wavefunctions,
respectively, in the relative coordinate between the paired fermions of the
CP. In Refs.\cite{BF7a}\cite{PLA2} they are taken simply as%
\begin{equation}
f_{+}(\epsilon )=\left\{ 
\begin{array}{cc}
f & \quad \text{for}\,\,E_{f}<\epsilon <E_{f}+\delta \varepsilon ,\quad  \\ 
0 & \text{otherwise,}%
\end{array}%
\right.   \label{f+}
\end{equation}%
\begin{equation}
f_{-}(\epsilon )=\left\{ 
\begin{array}{cc}
f & \quad \text{for}\,\,E_{f}-\delta \varepsilon <\epsilon <E_{f},\quad  \\ 
0 & \text{otherwise.}%
\end{array}%
\right.   \label{f-}
\end{equation}%
The quantities $E_{f}$ and $\delta \varepsilon $ are \textit{new}
phenomenological dynamical energy parameters (in addition to the positive BF
vertex coupling parameter $f$) that replace the previous such $E_{\pm }(0)$,
through the relations 
\begin{equation}
E_{f}\equiv \tfrac{1}{4}[E_{+}(0)+E_{-}(0)]\text{ \ \ \ and \ \ \ }\delta
\varepsilon \,\equiv \,\tfrac{1}{2}[E_{+}(0)-E_{-}(0)],  \label{27}
\end{equation}%
where $E_{\pm }(0)$ are the (empirically \textit{un}known) zero-CMM energies
of the 2e- and 2h-CPs, respectively. Clearly $E_{\pm }(0)=2E_{f}\pm \delta
\varepsilon $. The quantity\ $E_{f}$ will serve as a convenient energy scale
and is not to be confused with the Fermi energy $E_{F}=\tfrac{1}{2}%
mv_{F}^{2}\equiv k_{B}T_{F}$ where $T_{F}$\ is the Fermi temperature. The
Fermi energy $E_{F}$ equals $\pi \hbar ^{2}n/m$ in 2D and $(\hbar
^{2}/2m)(3\pi ^{2}n)^{2/3}$ in 3D, with $n$ the total number-density of
charge-carrier electrons. The quantities $E_{f}$ and $E_{F}$ coincide 
\textit{only }when perfect 2e/2h-CP symmetry holds.

The interaction Hamiltonian (\ref{Hint}) can be further simplified by
keeping only the $\mathbf{K}=0$ terms. One can then apply the Bogoliubov
``recipe'' \cite{Bog47}\ (see, also e.g., \cite{FW} p. 199) of replacing in
the full Hamiltonian $H=H_{0}+H_{int}$ all zero-CMM 2e- and 2h-CP boson
creation and annihilation operators by their respective c-numbers $\sqrt{%
N_{0}}$ and $\sqrt{M_{0}}$, where $N_{0}(T)$ and $M_{0}(T)$ are the number
of zero-CMM 2e- and 2h-CPs, respectively. One eventually seeks the lowest
critical temperature $T_{c}$\ such\ that, e.g., $N_{0}(T_{c})$ or $%
M_{0}(T_{c})$ vanish. Note that $T_{c}$ calculated thusly can, in principle,
turn out to be zero, in which case there is no BEC. Now, $H-\mu \hat{N}$ can
be diagonalized exactly via a Bogoliubov-Valatin transformation \cite{Bog}\
in terms of the thermodynamic potential $\Omega \equiv -PL^{d}$\ for the
CBFM, with $P$ its pressure and $L^{d}$ the system ``volume,'' which is%
\begin{equation}
\Omega (T,L^{d},\mu ,N_{0},M_{0})\text{\ }=-k_{B}T\ln \left[ \text{Tr}%
e^{-\beta (H-\mu \hat{N})}\right] ,  \label{28}
\end{equation}%
where ``Tr'' stands for ``trace.'' Inserting (\ref{H0}) and (\ref{Hint})
into (\ref{28}) one obtains \cite{BF7a}\ after some algebra 
\begin{gather}
\Omega (T,\,L^{d},\text{ }\mu ,\,N_{0},\,M_{0})/L^{d}=  \notag \\[0.05in]
=\int_{0}^{\infty }d\epsilon N(\epsilon )\,[\epsilon -\mu -E(\epsilon
)]-2\,k_{B}T\int_{0}^{\infty }d\epsilon \,N(\epsilon )\,\ln \{1+\exp [-\beta
\,E(\epsilon )]\}+  \notag \\
+[E_{+}(0)-2\,\mu ]\,n_{0}+k_{B}T\int_{0+}^{\infty }d\eta \,M(\eta )\,\ln
\{1-\exp [-\beta \mathcal{E}_{+}(\eta )]\}+  \notag \\
+[2\,\mu -E_{-}(0)]\,m_{0}+k_{B}T\int_{0+}^{\infty }d\eta \,M(\eta )\,\ln
\{1-\exp [-\beta \mathcal{E}_{-}(\eta )]\}.  \label{29}
\end{gather}%
For $d=3$ one has%
\begin{equation}
N(\epsilon )\equiv \text{ \ }\frac{m^{3/2}}{2^{1/2}\pi ^{2}\hbar ^{3}\,}%
\sqrt{\,\epsilon }\text{ \ \ and \ \ }M(\eta )\equiv {\Huge \,}\frac{2m^{3/2}%
}{\pi ^{2}\hbar ^{3}}\sqrt{\eta }  \label{30}
\end{equation}%
for the (one-spin) fermion and boson DOS at energies $\epsilon =\hbar
^{2}k^{2}/2m$ and $\eta =$ $\hbar ^{2}K^{2}/2(2m)$, respectively. The latter
is an \textit{assumption }to be lifted later so as to include Fermi sea
effects which transform the boson dispersion relation from quadratic to
linear, as already discussed in Sections 3 to 5. Also, recalling the 2e- and
2h-boson energies $E_{\pm }(K)$\ introduced in (\ref{H0}) the (bosonic)
energies $\mathcal{E}_{+}(\eta )$ and $\mathcal{E}_{-}(\eta )$ to simplify
notation have been defined as%
\begin{gather}
E_{+}(K)-2\mu =E_{+}(0)-2\mu +\eta \,\equiv \text{ }\mathcal{E}_{+}(\eta ),
\label{31} \\
2\mu -E_{-}(K)=2\mu -E_{-}(0)+\eta \equiv \text{ }\mathcal{E}_{-}(\eta ).
\label{32}
\end{gather}%
Finally, the relation between the fermion spectrum $E(\epsilon )$ and
fermion energy gap $\Delta (\epsilon )$ is of the form%
\begin{gather}
E(\epsilon )=\text{ }\sqrt{(\epsilon -\mu )^{2}+\Delta ^{2}(\epsilon )},
\label{33} \\
\Delta (\epsilon )\equiv \sqrt{n_{0}}f_{+}(\epsilon )+\sqrt{m_{0}}%
f_{-}(\epsilon ).  \label{34}
\end{gather}%
This last expression for the gap $\Delta (\epsilon )$ implies a simple $T$%
-dependence rooted in the 2e-CP $n_{0}(T)\equiv N_{0}(T)/L^{d}$ and 2h-CP $%
m_{0}(T)\equiv M_{0}(T)/L^{d}$ number densities of BE-condensed bosons, i.e.,%
\begin{equation}
\Delta (T)=\sqrt{n_{0}(T)}f_{+}(\epsilon )+\sqrt{m_{0}(T)}f_{-}(\epsilon ).
\label{35}
\end{equation}%
A $\Delta (T)^{2}$ temperature dependence in the order parameter in thin
films of the $Tl_{2}Ba_{2}CaCu_{2}O_{8}$ cuprate suggests itself from
thermal difference reflectance spectra \cite{Little}. This implies that
condensate densities $n_{0}(T)$ or $m_{0}(T)$ might be more fundamental as
order parameters than $\Delta (T)$, at least for this material.

Minimizing $\Omega (T,L^{d},\mu ,N_{0},M_{0})$ with respect to $N_{0}$ and $%
M_{0}$, and simultaneously fixing the total number $N$ of electrons by
introducing the electron chemical potential $\mu $,\ one thus specifies an 
\textit{equilibrium state} of system with volume $L^{d}$ and temperature $T$
by requiring that 
\begin{equation}
\frac{\partial \Omega }{\partial N_{0}}\text{ \ }=\text{ \ }0,\,\text{\ \ \
\ \ \ }\frac{\partial \Omega }{\partial M_{0}}\text{\ \ }=\text{ \ }0,\text{
\ \ \ \ \ \ \ and \ \ \ \ \ \ \ \ }\frac{\partial \Omega }{\partial \mu }%
\text{ \ }=\text{ \ }-N.  \label{36}
\end{equation}%
Here $N$ evidently includes both paired and unpaired CP fermions. Some
algebra then leads to the three coupled transcendental Eqs. (7)-(9) of Ref. %
\cite{BF7a}. These can be rewritten somewhat more transparently as: a) two 
\textit{``gap-like equations''}%
\begin{equation}
2\sqrt{n_{0}}[E_{+}(0)-2\mu ]=\int\limits_{0}^{\;\infty }d\epsilon
N(\epsilon )\frac{\Delta (\epsilon )f_{+}(\epsilon )}{E(\epsilon )}\tanh 
{\frac12}%
\beta E(\epsilon );  \label{37}
\end{equation}%
\begin{equation}
2\sqrt{m_{0}}[2\mu -E_{-}(0)]=\int\limits_{0}^{\;\infty }d\epsilon
N(\epsilon )\frac{\Delta (\epsilon )f_{-}(\epsilon )}{E(\epsilon )}\tanh 
{\frac12}%
\beta E(\epsilon ),  \label{38}
\end{equation}%
and b) a single \textit{``number equation'' }(that ensures charge
conservation)%
\begin{equation}
2n_{B}(T)-2m_{B}(T)+n_{f}(T)=n.  \label{39}
\end{equation}%
Here $n\equiv N/L^{3}$ is the number density of electrons while\ $n_{B}(T)$
and $m_{B}(T)$ are the number densities of 2e- and 2h-CPs in \textit{all }%
bosonic states, respectively, and are given by%
\begin{eqnarray}
\text{\ }n_{B}(T) &\equiv &n_{0}(T)\text{ }+\text{ }\int\limits_{0+}^{\infty
}d\eta M(\eta )\frac{1}{e^{\beta \mathcal{E}_{+}(\eta )}-1},{\Huge \!}
\label{40} \\
m_{B}(T) &\equiv &m_{0}(T)+\int\limits_{0+}^{\infty }d\eta M(\eta )\frac{1}{%
e^{\beta \mathcal{E}_{-}(\eta )}-1}.  \label{41}
\end{eqnarray}%
Clearly, 
\begin{equation}
n_{f}(T)\equiv \int\limits_{0}^{\;\infty }d\epsilon N(\epsilon )[1-\frac{%
\epsilon -\mu }{E(\epsilon )}\tanh 
{\frac12}%
\beta E(\epsilon )]  \label{42}
\end{equation}%
is the number density of unpaired electrons at any $T$. Self-consistent (at
worst, numerical) solution of the \textit{three coupled equations}\textbf{\ }%
(\ref{37}) to (\ref{39})\textbf{\ }yields the three thermodynamic variables
of the CBFM%
\begin{equation}
\ \ n_{0}(T,n,\mu ),\;\ \ \ m_{0}(T,n,\mu ),\ \ \ \text{and}\ \ \ \mu (T,n).
\label{43}
\end{equation}

The pressure $P$, entropy $S$ and specific heat at constant volume $C$ of an
equilibrium state characterized by $T$ and $n$ are then given by 
\begin{gather}
P(T,n)=-\Omega /L^{d},\;\ \text{\ \ }S(T,n)/L^{d}=-k_{B}\frac{\partial }{%
\partial T}(\Omega /L^{d}),\;\text{\ \ }  \label{44} \\
\quad C(T,n)/L^{d}=T\frac{\partial }{\partial T}\left[ S(T,n)/L^{d}\right] ,
\label{45}
\end{gather}%
all evaluated at fixed $n_{0}(T,\mu ,n)$, $m_{0}(T,\mu ,n)$ and $\mu (T,n)$.
The Helmholtz free energy 
\begin{equation}
F(T,L^{d},N)\equiv E-TS\equiv \Omega +\mu N  \label{45a}
\end{equation}
where $E$\ is the internal energy,\ then follows from 
\begin{equation}
F(T,n)/L^{d}=-P(T,n)+n\,\mu (T,n).  \label{46}
\end{equation}%
Finally, the critical magnetic field is 
\begin{gather}
H_{c}^{2}(T,n)/8\,\pi \equiv F_{n}(T,n)/L^{d}-F_{s}(T,n)/L^{d}=  \notag \\
=P_{s}(T,n)-P_{n}(T,n)+[\mu _{n}(T,n)-\mu _{s}(T,n)]\,n,  \label{47}
\end{gather}%
with subscripts $s$ and $n$ meaning ``superconducting'' and ``normal''
states.

\section{Main statistical theories as special cases of CBFM}

Most significantly, the three CBFM equations just stated contain \textit{%
five }different theories as special cases, see flow chart in Fig. 7. Perfect
2e/2h CP symmetry $n_{B}(T)=m_{B}(T)$ \textit{and }$n_{0}(T)=m_{0}(T)$ can
be seen from (\ref{37}) and (\ref{38}), as well as from $E_{\pm
}(0)=2E_{f}\pm \delta \varepsilon $, to imply that $E_{f}$ coincides with $%
\mu $. The CBFM then reduces to:

\qquad \textbf{i)} the gap and number equations of the \textit{BCS-Bose
crossover picture} \cite{BCS-Bose} for the BCS model interaction---if the
BCS parameters $V$ and Debye energy $\hbar \omega _{D}$ are identified with
the BF interaction Hamiltonian $H_{int}$ parameters $f^{2}/2\delta
\varepsilon $ and $\delta \varepsilon $, respectively. The crossover picture
for unknowns $\Delta (T)$ and $\mu (T)$ is now supplemented by the key
relation 
\begin{equation}
\Delta (T)=f\sqrt{n_{0}(T)}=f\sqrt{m_{0}(T)}.  \label{sqrt}
\end{equation}%
The 35-year-old crossover picture is associated with names such as (in
chronological order) Friedel and co-workers \cite{Friedel}, Eagles \cite%
{Eagles}, Leggett \cite{Leggett}, Miyake \cite{Miyake}, Nozieres \cite%
{Nozieres}, Micnas, Ranninger \& Robaszkiewicz \cite{BF3}, Randeria \cite%
{Randeria89}, van der Marel \cite{vanderMarel}, Bar-Yam \cite{BF6aa},
Drechsler \& Zwerger \cite{DZ92}, Hausmann \cite{Haus}, Pistolesi \&
Strinati \cite{Pistolesi}, and many others. However, it seems to be a very
modest improvement over BCS theory since the unphysically large $\lambda $
of about 8 is required to bring $\mu (T_{c})/E_{F}$ down from 1.00 to 0.998 %
\cite{FJ}. If one imposes that $\mu (T_{c})=E_{F}$ exactly, as follows for
weak BF coupling $f$\ from the number equation,\ the crossover picture is
well-known to reduce to:

\qquad \textbf{ii)} \textit{ordinary BCS theory }which is characterized by a 
\textit{single} equation, the gap equation for all $T$. Thus, \textit{the
BCS condensate is precisely a BE condensate} whenever both $n_{B}(T)=m_{B}(T)
$ and $n_{0}(T)=m_{0}(T)$ \textit{and }the BF coupling $f$ is small. Indeed,
for small coupling $\lambda $ the CBFM $T=0$ superconducting state has a 
\textit{lower }energy than the corresponding BCS state, see Appendix B. The
BCS state comes from a variational trial wave function so that its energy
expectation value by the Rayleigh-Ritz variational principle is a rigorous
upper bound to the exact value. Thus, the CBFM has a somewhat larger
condensation energy than that of BCS theory, but both \textit{coincide} in
leading order. In addition, at least two universal constants, the gap-to-$%
T_{c}$ ratio $2\Delta (0)/k_{B}T_{c}$ and the specific heat jump $\Delta
C_{s}(T_{c})$ to the normal value $C_{n}(T_{c})$, have been shown \cite{PLA2}%
\ to coincide with the BCS values\ of $3.53$ and $1.43$ in this limit of the
CBFM.

On the other hand, for no 2h-CPs present the CBFM reduces \cite{BF7a}\ also
to:

\qquad \textbf{iii)} the \textit{BEC BF model} in 3D of Friedberg and Lee %
\cite{BF5,BF6} characterized by the relation $\Delta (T)=f\sqrt{n_{0}(T)}$.
With just \textit{one} adjustable parameter (the ratio of perpendicular to
planar boson masses) this theory fitted \cite{BF6}\ cuprate $T_{c}/T_{F}$
data quite well. The ratio turned out to be 66,560---just under the $10^{5}$
anisotropy ratio reported \cite{Fiory} almost contemporaneously for $BSCO.$\
Finally, when $f=0$ this model reduces\ to:

\qquad \textbf{iv)} the ideal BF model (IBFM) of Ref. \cite{BF9,BF10} that
predicts nonzero BEC $T_{c}$s even in 2D. The ``gapless'' IBFM\ cannot
describe the superconducting phase. But considered as a model for the 
\textit{normal state} it should provide feasible $T_{c}$s as singularities
within a BE scenario that are approached from \textit{above }$T_{c}$. Figure
8 displays the $T_{c}$ predictions \cite{BF10} in 2D for cuprate
superconductors with \textit{no} adjustable parameters. Finally, the CBFM
reduces to:

\qquad \textbf{v)} the familiar $T_{c}$-formula of ordinary BEC in 3D,
albeit as an \textit{implicit }equation with the boson number-density being $%
T$-dependent. 

\section{Enhanced $T_{c}$s from the CBFM}

The very general CBFM has been applied and gives sizeable enhancements in $%
T_{c}$s over BCS theory that emerge for moderate departures from perfect
2e/2h-pair symmetry. This is attained for the \textit{same }interaction
model (with the same coupling strength $\lambda $ and cutoff $\hbar \omega
_{D}$\ parameters) used in conventional superconductors. The BF
(two-fermion) interaction (\ref{Hint}) with (\ref{f+}) and (\ref{f-}) bears
a one-to-one correspondence with the more familiar ``direct'' four-fermion
electron-phonon interaction, mimicked, e.g., by the BCS model interaction.
Its double Fourier transform is a negative constant $-V$, nonzero only
within an energy shell $2\hbar \omega _{D}$ about the Fermi surface, with $%
\omega _{D}$ the Debye frequency. The correspondence is realized \cite%
{BF7a,PLA2} by setting $f^{\,2}/2\,\delta \varepsilon \equiv V$ and $\delta
\varepsilon \equiv \hbar \,\omega _{D}$, from which the familiar
dimensionless BCS model interaction parameters $\lambda \equiv N(E_{F})V$
and $\hbar \,\omega _{D}/E_{F}$ are then recovered.

The three coupled equations (\ref{37}) to (\ref{39}) of the CBFM\ that
determine the $d$-dimensional BE-condensate number-densities $n_{0}(T)$ and $%
m_{0}(T)$ of 2e- and 2h-CPs, respectively, as well as the fermion chemical
potential $\mu (T)$, were solved numerically in 3D for $\lambda =1/5$ and $%
\hbar \omega _{D}/E_{F}=$ $0.001$ in Ref. \cite{PLA2} assuming a quadratic
boson dispersion relation $\eta =\hbar ^{2}K^{2}/2(2m)$, i.e., $s=2$ in (\ref%
{0}).\ For this case Figure 9 maps the phase diagram in the vicinity of the
BCS $T_{c}$ value (marked BCS-B in the figure) at $\Delta n\equiv n/n_{f}-1=0
$ (corresponding to perfect 2e/2h-CP symmetry) where $n_{f}$ is a very
special case of $n_{f}(T)$\ as defined in (\ref{42}). Namely, $n_{f}$ can be
seen \cite{PLA2} to be the number of unpaired electrons at zero coupling and
temperature. Besides the \textit{normal }phase ($n$) consisting of the ideal
BF gas described by $H_{0}$, three different stable BEC phases
emerged---plus several metastable, i.e., of higher Helmholtz free energy.
These are two \textit{pure} phases of \textit{either} 2e-CP ($s+$)\ \textit{%
or} 2h-CP ($s-$)\ BE-condensates, and a lower temperature \textit{mixed }%
phase ($ss$) with arbitrary proportions of 2e- \textit{and} 2h-CPs, see Fig.
9. Of greater physical interest are the two higher-$T_{c}$ \textit{pure }%
phases so that we focus below only on them. For each pure phase at a
critical temperature we have \textit{either} $\Delta (T_{cs+})=f\sqrt{%
n_{0}(T_{cs+})}$ $\equiv 0$ or $\Delta (T_{cs-})=f\sqrt{m_{0}(T_{cs-})}%
\equiv 0$, where $\Delta (T)$ is the electronic (BCS-like) energy gap. Their
intersection gives the BCS $T_{c}$ value of $7.64\times 10^{-6}T_{F}$ that
also follows from the familiar BCS expression 
\begin{equation}
T_{c}/T_{F}\simeq 1.134(\hbar \omega _{D}/E_{F})\exp (-1/\lambda ).
\label{54'}
\end{equation}

We next focus on $s=1$ which, as we saw in Sections 3 to 5, occurs in the
leading term for ``ordinary'' CPs in a Fermi sea as well as for
``generalized'' CPs in a BCS state.\ For these, the boson energy $\eta $ in (%
\ref{29}) to be used later has leading terms in the many-body Bethe-Salpeter
(BS) CP dispersion relation that are%
\begin{equation}
\eta \simeq (\lambda /2\pi )\hbar v_{F}K\text{ \ \ \ \ \ \ \ \ \ \ \ \ \ \ 2D%
}  \label{50}
\end{equation}%
\cite{ANFdeLl} and 
\begin{equation}
\eta \simeq (\lambda /4)\hbar v_{F}K\text{ \ \ \ \ \ \ \ \ \ \ \ \ \ \ \ \ 3D%
}  \label{51}
\end{equation}%
\cite{Honolulu}. As before, $\lambda \equiv VN(E_{F})$ where $N(E_{F})$ is
the electron DOS (for one spin) at the Fermi surface. Note that the boson
energy $\eta $ is \textit{not }the quadratic $\hbar ^{2}K^{2}/2(2m)$
appropriate for a composite boson of mass $2m$ moving not in the Fermi sea
but in vacuum \cite{Blatt}-\cite{BF2}, \cite{BF5}-\cite{PLA2}, and assumed
in (\ref{30}). Note also that the BS CP linear dispersion coefficient $%
\lambda /2\pi $ in 2D (or $\lambda /4$\ in 3D) contrasts markedly with the
coupling-independent $2/\pi $ coefficient in 2D (or $1/2$ in 3D, as
apparently first quoted in Ref. \cite{Schrieffer}, p. 33) obtained \cite%
{PC98}\ in the \textit{simple} CP problem \cite{Coo} which ignores holes.
Thus, again with $n_{B}=n/2$, for $s=1$ and $C_{1}=\lambda b(d)\hbar v_{F}$
with $b(2)=1/2\pi $ and $b(3)=1/4$ according to (\ref{50}) and (\ref{51}), (%
\ref{C.1})\ leads to 
\begin{equation}
T_{c}/T_{F}=2\lambda b(d)/\left[ d\Gamma (d)\zeta (d)\right] ^{1/d}.
\label{52}
\end{equation}%
For $\lambda =1/2$ this is $\simeq 0.088$ if $d=2$ since $\zeta (2)=\pi
^{2}/6$, and $\simeq 0.129$ if $d=3$ since $\zeta (3)\simeq 1.202$.\ These
two values will appear as the black squares in Figs. 10 to 13 (upper ones in
Figs. 10 and 11). They mark the BEC limiting values if \textit{all }%
electrons\ in our 2D or 3D many-electron system were imagined paired into
noninteracting bosons formed with the BCS model interelectron interaction.
For $\lambda =1/4$ the lower black squares in Figs. 10 and 11 apply.

\subsection{Two dimensions (2D)}

To address cuprates and other copper-plane-free planar superconductors (such
as $Sr_{2}YRu_{1-x}Cu_{x}O_{6}$ with $T_{c}\simeq 49$ $K,$\ which is $Cu$%
-doped but has no $CuO$ planes)\ we deal first with the CBFM in 2D. Although
it is still controversial \cite{Dow}\ as to \textit{which} planes, the $CuO$
or $BaO$ or $SrO$, etc., the superconductivity resides in, these planes are
parallel to each other, say, in the $ab$ directions.\ Resistivity $\rho $
anisotropies $\rho _{c}/\rho _{ab}$, with the $c$ direction perpendicular to
the $ab$ denoting the $CuO$ or $BaO$ or $SrO$ planes, can be as high as $%
10^{5}$ in $Bi_{2+x}Sr_{2-y}CuO_{6+\delta }$ \cite{Fiory}, if not higher,
even though ``only'' about $10^{2}$ in $YBa_{2}Cu_{3}O_{7-\delta }$ $(YBCO)$%
. From the Drude 1900 resistivity model\ \cite{AandM}$\ \rho =m/ne^{2}\tau $
for current carriers of charge $e$, effective mass $m$, number density $n$,
and $\tau $ is some average time between collisions. Thus, if $\rho
_{c}/\rho _{ab}=m_{c}/m_{ab}$\ is $\infty $ one has a precisely 2D
situation; if it is $1$ we have the perfectly isotropic 3D case. Hence, the
large ($10^{5}$) but finite ratio observed implies $(2+\epsilon )D$ or
``quasi-2D.'' For cuprates the value $d\simeq 2.03$ has been extracted
independently by two groups \cite{wen}\cite{Panat} for $YBCO$ as more
realistic than $d=2,$ since that value reflects inter-CuO (or BaO)-layer
couplings. Our results in that case would be very close to those for $d=2$
since, e.g., from Appendix C (\ref{C.1}), $T_{c}$ for $s=1$ (but, perhaps
very significantly, \textit{not }for $s=2$)\ varies little\ \cite{Sevilla}\
with $d$ around $d=2$. This justifies models in exactly 2D, at least as a
very good initial approximation.

In 2D the electronic DOS per unit area $L^{2}$\ is constant, namely $%
N(\varepsilon )=m/2\pi \hbar ^{2}$. Using the leading term of the BS CP 
\textit{linear }dispersion relation (\ref{50}) we get for the bosonic DOS 
\begin{equation}
M(\eta )\equiv (1/2\pi )K(dK/d\eta )\simeq (2\pi /\lambda ^{2}\hbar
^{2}v_{F}^{2})\eta   \label{53}
\end{equation}%
instead of the constant that follows in 2D from the quadratic dispersion $%
\eta =$ $\hbar ^{2}K^{2}/2(2m)$. Employing $E_{f}\equiv \pi \hbar
^{2}n_{f}/m=k_{B}T_{f}$ as energy/density/temperature scaling factors, and
the relation $n/n_{f}=(E_{F}/E_{f})^{d/2}$ to convert quantities such as $%
T_{c}/T_{f}$ to $T_{c}/T_{F}$, where $E_{F}\equiv $ $k_{B}T_{F}$, the two
``working equations'' for the \textit{pure 2e-CP phase }[i.e., $%
m_{B}(T_{c})\equiv 0$] (with all quantities dimensionless, energies in units
of $E_{f}$ and electron particle-densities in units of $n_{f}$) become%
\begin{equation}
1+\hbar \omega _{D}/2-\mu =\lambda (\hbar \omega
_{D}/2)\int\limits_{1}^{1+\hbar \omega _{D}}dx\frac{1}{|x-\mu |}\mathrm{%
\tanh }\frac{|x-\mu |}{2T_{c}},  \label{mu1}
\end{equation}%
\begin{equation}
\frac{1}{2}\int\limits_{0}^{\infty }dx[1-\mathrm{\tanh }\frac{x-\mu }{2T_{c}}%
]+\frac{\pi ^{2}}{\lambda ^{2}}\frac{1}{n}\int\limits_{0}^{\infty }dxx[%
\mathrm{\coth }\,\frac{x+2(1+\hbar \omega _{D}/2-\mu )}{2T_{c}}-1]=n.
\label{n1}
\end{equation}%
These are just the gap-like equation associated with 2e-CPs and its
corresponding number equation. Recall that $n_{f}$ can be seen \cite{BF7a}
to be the number-density of unpaired electrons at zero $T$ and zero $\lambda 
$.

For the \textit{pure 2h-CP phase\ }(i.e., $n_{B}\equiv 0$) the two working
equations are%
\begin{equation}
\mu -1+\hbar \omega _{D}/2=\lambda (\hbar \omega _{D}/2)\int\limits_{1-\hbar
\omega _{D}}^{1}dx\frac{1}{|x-\mu |}\mathrm{\tanh }\frac{|x-\mu |}{2T_{c}},
\label{mu2}
\end{equation}

\begin{equation}
\frac{1}{2}\int\limits_{0}^{\infty }dx[1-\mathrm{\tanh }\frac{x-\mu }{2T_{c}}%
]-\frac{\pi ^{2}}{\lambda ^{2}}\frac{1}{n}\int\limits_{0}^{\infty }dxx[%
\mathrm{\coth }\frac{x-2(1-\hbar \omega _{D}/2-\mu )}{2T_{c}}-1]=n.
\label{n2}
\end{equation}%
which are the gap-like equation associated with 2h-CPs and its corresponding
number equation. Note that (\ref{n1}) and (\ref{n2}) are quadratic in $n$.
Further, all integrals there are exact, i.e.,%
\begin{equation}
\int\limits_{0}^{\infty }dx[1-\mathrm{\tanh }\frac{x-\mu }{2T_{c}}]=\mu
+2T_{c}\mathrm{ln}[2\mathrm{cosh}({\mu /2T}_{c})],  \label{mu3}
\end{equation}%
and%
\begin{equation}
\int\limits_{0}^{\infty }dxx[\mathrm{\coth }\frac{x+\delta }{2T_{c}}%
-1]=2T_{c}^{2}\mathrm{PolyLog[2,}\text{ }e^{-\delta /T_{c}}]  \label{n3}
\end{equation}%
where the polylogarithm function as defined in Ref. \cite{Wolfram} is 
\begin{equation}
PolyLog[\sigma ,z]\equiv \sum\limits_{\text{ }l\text{ }\mathbf{=1}}^{\infty
}z^{l}/l^{\sigma }.  \label{polylog}
\end{equation}%
The integrals in (\ref{mu1}) and (\ref{mu2}), however, were performed
numerically. In 2D we use the two extreme values of $\lambda =1/4$ (lower
set of curves in Figs. 10 and 11) and $=1/2$ (upper set of curves), and $%
\hbar \omega _{D}/E_{F}=0.05$ (a typical value for cuprates). If $\lambda
>1/2$ the ionic lattice in 3D becomes unstable, \cite{Migdal} and Ref. \cite%
{Blatt} p. 204. Equations (\ref{mu1}) to (\ref{n2}) lead to the $T_{c}/T_{F}$
\textit{vs.} $n/n_{f}$ phase diagram which is graphed in Figs. 10 and 11 for
both 2e-CP (dashed curve) and 2h-CP (full curve) pure, stable BEC-like
phases. The value $n/n_{f}=1$ corresponds to perfect 2e/2h-CP symmetry for
the lower-$T_{c}$ mixed phase (not shown), while $n/n_{f}>1$ (and $<1)$
signifies more (less) 2e-CPs than 2h-CPs in the mixed phase. The $T_{c}$
value where both phase-boundary curves $n_{0}(T_{c})=m_{0}(T_{c})$ $=0$
intersect\ is marked by the large dots in the figure. The values are
consistent with those gotten from (\ref{54'}) which gives $\simeq 0.001$ for 
$\lambda =1/4$ and $\simeq 0.008$ for $\lambda =1/2$, for $\hbar \omega
_{D}/E_{F}=0.05$. [The values from (\ref{54'}) hardly differ from those from
the exact BCS (implicit) $T_{c}$-formula (Ref. \cite{FW}, p. 447) $1$ $%
=\lambda \int_{0}^{\hbar \omega _{D}/2k_{B}T_{c}}dxx^{-1}\tanh x$.] Cuprate
data empirically \cite{Poole} show $T_{c}$'s\ and $T_{F}$'s\ falling within
the range $T_{c}/T_{F}\simeq 0.03-0.09$. Thus, moderate departures from
perfect 2e/2h-CP symmetry enable the CBFM to predict quasi-2D cuprate
empirical $T_{c}$ values, \textit{without abandoning electron-phonon dynamics%
}---contrary to popular belief.

As the variable $n_{f}$ used to scale the electron number density $n$ in
Fig. 10 does not appear at all in the number equation (\ref{39}), we also
scaled $n$ with the variable $n_{f}(T)$ evaluated at $T=T_{c}$\ which does
appear in (\ref{39}) and which has a clear-cut physical meaning. Since $%
\Delta (T_{c})=0$, from (\ref{33}) and (\ref{34}), (\ref{42}) reduces in 2D
to the analytic expression 
\begin{equation}
n_{f}(T_{c})/n\equiv (T_{c}/T_{F})\ln [1+\exp \{\mu (T_{c})/k_{B}T_{c}\}].
\label{55}
\end{equation}%
For the pure 2e-CP condensate case, i.e., $m_{B}(T)\equiv 0,$\ we must have $%
n/n_{f}(T_{c})\geq 1$ from (\ref{39}), where in general $n_{f}(T_{c})\neq
n_{f}$. Here $n_{f}(T_{c})$\ is the actual number density of unpaired
electrons at $T=T_{c}$ since the number equation (\ref{39}) for $m_{B}(T)=0$
then becomes just 
\begin{equation}
n_{f}(T_{c})+2n_{B+}(T_{c})=n.  \label{56}
\end{equation}%
Here $n_{B+}(T_{c})$ is the total number density of all \textit{excited} (or 
$K>0$) 2e-CPs, also commonly known as ``preformed CPs.'' It is defined in (%
\ref{40}) at any $T$ as%
\begin{equation}
n_{B+}(T)\equiv \int\limits_{0^{+}}^{\infty }d\eta M(\eta )\frac{1}{e^{\beta 
\mathcal{E}_{+}(\eta )}-1}.  \label{57}
\end{equation}%
Preformed CPs seem to be the main conjecture to explain a ``pseudogap'' \cite%
{Timusk99} observed above $T_{c}$ in many underdoped cuprates.\ 

The complete number equation (\ref{39}) then explicitly reads%
\begin{equation}
2n_{0}(T)+2n_{B+}(T)-2m_{0}(T)-2m_{B+}(T)+n_{f}(T)=n  \label{57'}
\end{equation}%
with $m_{B+}(T)$ being the number of preformed 2h-CPs, defined as the
integral term in (\ref{41}).\ Using $n_{f}(T_{c})$ instead of $n_{f}$\ to
scale electron densities $n$ in (\ref{mu1}) to (\ref{n2}) which are solved
numerically, scaling all energies with $E_{F}$ instead of $E_{f}$ as before,
and on eliminating $n_{f}$, gives the results in Fig. 11. Curiously, these
curves are hard to distinguish from those of Fig. 10. In Fig. 11 the BCS
values (black dots in figure) are \textit{not }associated as before with the
intersection of the 2e- and 2h-CP condensation curves, but are merely values
calculated with the BCS $T_{c}$ formula (\ref{54'}).\ On the other hand, the 
$T_{c}$ values from the CBFM equations scaled with the previous (and
somewhat opaque) variable $n_{f}$, match only \textit{approximately }in the
limit $n/n_{f}\rightarrow \infty $\ with the analytical BEC limits of (\ref%
{52}). It has been verified numerically that coincidence in the BEC limit is
perfect only in the trivial limit of zero coupling, as expected.

In view of Figs. 10 and 11 we note that \textit{room temperature
superconductivity }(RTSC in figures) is possible, e.g., in Fig. 11, for $%
n/n_{f}(T_{c})$ sufficiently less than unity. From (\ref{57'}) with $%
n_{B}(T)\equiv n_{0}(T)+n_{B+}(T)\equiv 0$ this translates into this recipe: 
$2m_{B+}(T_{c})$ must be sufficiently larger than $n_{f}(T_{c})$.
Specifically, from Fig. 11 RTSC occurs for $\lambda =%
{\frac12}%
$ if $n/n_{f}(T_{c})\lesssim 0.1$, which from (\ref{39}) means\ $%
2m_{B+}(T_{c})\gtrsim 0.9n_{f}(T_{c})$ or that the total number of
individual holes bound into \textit{preformed} 2h-CPs at $T_{c}$ must
somehow be larger than about 90\% of the number of unpaired electrons at $%
T_{c}$.

\subsection{Three dimensions (3D)}

In 3D the electronic DOS if $\epsilon \equiv \hbar ^{2}k^{2}/2m$ is the
familiar expression 
\begin{equation}
N(\epsilon )\equiv (1/2\pi ^{2})k^{2}(dk/d\epsilon )=(m^{\frac{3}{2}}/2^{%
\frac{1}{2}}\pi ^{2}\hbar ^{3})\sqrt{\epsilon }  \label{58}
\end{equation}%
which coincides with (\ref{30}). Analogously as before, $E_{f}=(\hbar
^{2}/2m)(3\pi ^{2}n_{f})^{2/3}\equiv k_{B}T_{f}$ again differs from $%
E_{F}=(\hbar ^{2}/2m)(3\pi ^{2}n)^{2/3}\equiv k_{B}T_{F}$ unless perfect
2e/2h-CP symmetry holds in which case they coincide. On the other hand, the
leading term in the BS CP boson dispersion energy is now the linear
expression (\ref{51}) so that 
\begin{equation}
M(\eta )\equiv (1/2\pi ^{2})K^{2}(dK/d\eta )\simeq (32/\pi ^{2}\lambda
^{3}\hbar ^{3}v_{F}^{3})\eta ^{2}.  \label{59}
\end{equation}%
The above working equations for the \textit{pure 2e-CP phase} in 3D (all
quantities dimensionless as before in terms of $E_{f}$ and $n_{f}$) are%
\begin{equation}
1+\hbar \omega _{D}/2-\mu =\lambda (\hbar \omega _{D}/2)\frac{1}{n^{1/3}}%
\int\limits_{1}^{1+\hbar \omega _{D}}dx\sqrt{x}\frac{1}{|x-\mu |}\mathrm{%
\tanh }\frac{|x-\mu |}{2T_{c}},  \label{mu3De}
\end{equation}%
\begin{equation}
\frac{3}{4}\int\limits_{0}^{\infty }dx\sqrt{x}[1-\mathrm{\tanh }\frac{x-\mu 
}{2T_{c}}]+\frac{12}{\lambda ^{3}}\frac{1}{n}\int\limits_{0}^{\infty
}dxx^{2}[\mathrm{\coth }\frac{x+2(1+\hbar \omega _{D}/2-\mu )}{2T_{c}}-1]=n,
\label{n3De}
\end{equation}%
while for the\textit{\ pure 2h-CP phase} they are%
\begin{equation}
\mu -1+\hbar \omega _{D}/2=\lambda (\hbar \omega _{D}/2)\frac{1}{n^{1/3}}%
\int\limits_{1-\hbar \omega _{D}}^{1}dx\sqrt{x}\frac{1}{|x-\mu |}\mathrm{%
\tanh }\frac{|x-\mu |}{2T_{c}},  \label{mu3Dh}
\end{equation}

\begin{equation}
\frac{3}{4}\int\limits_{0}^{\infty }dx\sqrt{x}[1-\mathrm{\tanh }\frac{x-\mu 
}{2T_{c}}]-\frac{12}{\lambda ^{3}}\frac{1}{n}\int\limits_{0}^{\infty
}dxx^{2}[\mathrm{\coth }\frac{x-2(1-\hbar \omega _{D}/2-\mu )}{2T_{c}}-1]=n.
\label{n3Dh}
\end{equation}%
Results in 3D are reported only for $\lambda =1/2$, and $\hbar \omega
_{D}/E_{F}=0.005$ was used here. In Figs. 12 and 13 the full curves are
again the 2h-CP BEC phase boundaries while the dashed curves are the 2e-CP
BEC ones. The large dot again marks the BCS $T_{c}/T_{F}$ values obtainable
from (\ref{54'})\ of $0.0001$ for $\lambda =1/4$ and of $0.0008$ for $%
\lambda =1/2$. Besides Ref. \cite{Poole},\ mpirical data for both exotic and
conventional, elemental superconductors in 3D are also graphed in Ref. \cite%
{Uemura}. We see that whereas BCS theory roughly reproduces $T_{c}/T_{F}$
values\ well for the latter SCs, it takes moderate departures from perfect
2e/2h-CP symmetry to access 3D exotic SC $T_{c}/T_{F}$ values, which
empirically fall within the range $0.01-0.1$. This is much larger than the
range $\lesssim 0.001$ for conventional (elemental) superconductors, also
shaded in the figure.

\section{Hole superconductivity}

Finally, we address the unique but mysterious role played by \textit{hole }%
charge carriers \cite{Chapnik}\ in the normal state of superconductors in
general. For example: a) of the cuprates, those that are hole-doped have
transition temperatures $T_{c}$ about \textit{six }times higher than
electron-doped ones; and even in conventional superconductors \cite{Hirsch}
b) over 80\% of all superconducting elements have positive Hall coefficients
(meaning hole charge carriers); while c) over 90\% of non-superconducting
metallic, non-magnetic elements have electron charge carriers. This greater
``efficiency'' of individual, unpaired hole carriers in producing higher $%
T_{c}$s is clearly reflected in Figs. 10 and 11 for both 2D and 3D
superconductors, at least insofar as pure 2h-CP BE condensates exhibiting
higher $T_{c}$s than those associated with pure 2e-CP BE condensates.

\section{Conclusions}

This review began with a survey of ``ordinary'' and ``generalized'' Cooper
pairing, and stressed that if hole pairing is treated on an equal footing
with electron pairing the original ``ordinary'' Cooper problem (based on the
pure-kinetic-energy unperturbed Hamiltonian) is meaningless. However,
``generalized'' Cooper pairs defined in terms of the Bethe-Salpeter equation
including both electron and hole pairs, in the ladder approximation, but
based on the BCS ground state as unperturbed Hamiltonian instead of on the
ideal Fermi gas sea, gives rise to physically meaningful positive-energy
resonances with a finite lifetime for CMM $K>0$ and infinite lifetime for $%
K=0$.

It was then sketched how five statistical continuum theories of
superconductivity---including both the BCS and BEC theories---are contained
as special limiting cases within a single theory, the ``complete
boson-fermion model'' (CBFM). This model includes, for the first time, both
two-electron and two-hole pairs in freely variable proportions, along with
unpaired electrons. Thus, the BCS and BEC theories are fully \textit{unified 
}within the CBFM. The BCS condensate [specified by a single equation, namely
the $T$-dependent gap equation for $\Delta (T)$] follows directly from the
CBFM \textit{as a BE condensate} through the condition for phase equilibria
when both total 2e- and 2h-pair number, as well as their condensate,
densities are \textit{equal} at a given temperature and coupling---provided
the coupling is weak enough so that the electron chemical potential $\mu $
roughly equals the Fermi energy $E_{F}$. Ordinary BEC theory, on the other
hand, is recovered from the CBFM when hole pairs are completely neglected,
the BF coupling $f$ is made to vanish, and the limit of zero unpaired
electrons is taken.

Lastly, the practical outcome of the BCS-BEC unification via the CBFM is
fourfold: a) \textit{enhancements} in $T_{c}$, by more than an
order-of-magnitude in 2D, and more than two orders-of-magnitude in 3D, are
obtained with pure electron-pair, and even more so with pure hole-pair, BE
condensates for the same electron-phonon dynamics mimicked by the BCS model
interaction; b) these enhancements in $T_{c}$ fall within empirical ranges
for 2D and 3D ``exotic'' SCs, whereas BCS $T_{c}$ values remain low and
within the empirical ranges for conventional, elemental SCs using standard
interaction-parameter values; c) hole-doped SCs are predicted in both 2D and
3D to have higher $T_{c}$'s than electron-doped ones, in agreement with
observation; and finally that d) room temperature superconductivity is
possible, with the\textit{\ same} interaction parameters\textit{\ }used in
BCS theory for conventional SCs,\ but only via hole-pair BE
condensates.\bigskip 

\textbf{Acknowledgements }I thank V.C. Aguilera-Navarro, J. Batle, M. Casas,
J.R. Clem, M. Fortes, F.J. Sevilla, M.A. Sol\'{\i}s, S. Tapia, V.V.
Tolmachev, O. Rojo, J.J. Valencia, A.A. Valladares and H. Vucetich for
discussions and/or for providing material prior to its publication. MdeLl
acknowledges UNAM-DGAPA-PAPIIT (Mexico) grant \# IN106401, and CONACyT
(Mexico) grant \# 41302, for partial support.

3. Wavefunction Feynman diagrams for 2p ($\psi _{+}$), 2h ($\psi _{-}$) and
ph ($\psi _{0}$) bound states arising from the BS equations. Shaded
rectangles designate diagrams that do not contribute in the IFG-based case.

4. a) Exact ``moving''\ 2p-CP (real) energy $\mathcal{E}_{K}$\ (in units of $%
E_{F}$) in 2D from (\ref{mCP})\ (full curves), compared with its linear
leading term (thin short-dashed lines) and its linear plus quadratic
expansion (long-dashed curves) both from (\ref{linquadmCP}), \textit{vs.}
CMM wavenumber $K$ (in units of $k_{F})$, for interaction (\ref{int})
parameters $\lambda =
{\frac14}
$ (lower set of curves) and $%
{\frac12}
$ (upper set of curves), and $\hbar \omega _{D}/E_{F}=0.05$. For reference,
leading linear term (\ref{22}) of trivial ABH sound mode is also plotted
(lower thick dashed line). b) 2p-CP lifetime as defined in (\ref{25}). c)
Analogy of ordinary and BCS-based-BS 2p-CPs with various confined states in
a 3D potential problem.

5. BE condensate-fraction curves $1-(T/T_{c})^{d/s}$ for bosons in $d=3$, $2$%
, or $1$ with dispersion relation (\ref{0}) with $s=2$ or $1,$ for a \textit{%
pure }phase of either 2e- or 2h-CPs as discussed in text, compared to
empirical data for 3D SCs ($Nb/Cu$ and $Sn$), two quasi-2D SCs ($Y123$ and $%
Bi2212$) and a quasi-1D SC ($4$ $A$-wide nanotubes). Data for the latter are
for $\Delta (T)/\Delta (0)$ but are plotted as $[\Delta (T)/\Delta (0)]^{2}$
so as to reflect 2h-CP condensate fraction $m_{0}(T)/m_{0}(0)$\ according to
(\ref{sqrt}). The curve marked $1/1$ strictly corresponds to $T_{c}\equiv 0$%
; however, it serves as a lower bound to all curves with $d/s=(1+\epsilon )/1
$ for small but nonzero $\epsilon $ for which $T_{c}>0.$ The ordinate axis
is labeled with the superelectron number density $n_{s}(T)$ in units of the
normal electron density $n$. Also shown for reference are the two-fluid
model \cite{GorterCasimir}\ curve\ $1-(T/T_{c})^{4}$ and the BCS gap $\Delta
(T)/\Delta (0)$ order parameter curve \cite{Muhlsch}.

6. BF vertex diagrams depicting disintegration of 2e-CPs into two unpaired
electrons; 2e-CP formation process from the latter; formation of a 2h-CP
from two unpaired holes; and the disintegration of a 2h-CP into the latter.
All four processes are contained in $H_{int}$ Eq. (\ref{Hint}).

7. Flow chart of how the CBFM reduces in special cases to the statistical,
continuum models of superconductivity discussed in text, thereby displaying
how both BCS and BEC theories are unified. The three ``legs'' refer to:
perfect 2e-/2h-CP symmetry (middle leg, giving rise to the BCS-Bose
crossover and BCS theories); to no 2h-CPs present (right leg) leading to
ordinary BEC as a special case; and to no 2e-CPs present (left leg)
predicting room temperature superconductivity (RTSC), see Figs. 10 and 13
below, via a 2h-CP BE condensate.

8. Critical 2D BEC-like temperature ($T_{c}$ in units of $T_{F})$ approached
from \textit{above} within the ideal BF model (IBFM) with the BCS model
interaction for $\lambda =1/2$ for varying $\hbar \omega _{D}/E_{F}\equiv
\Theta _{D}/T_{F}$. Results are for: the pure unbreakable-boson gas with 
\textit{some} and with \textit{all} fermions paired; for the breakable-boson
gas; and for the boson-fermion mixture in thermal/chemical equilibrium
(thick full curve labeled ``binary gas''), all as described in Ref. \cite%
{BF10} for original (simple) CPs where $C_{1}=(2/\pi )\hbar v_{F}$ in (\ref%
{0}). Dashed curve is the BCS theory $T_{c}$, given by (\ref{54'}). Cuprate
experimental data are taken from Ref. \cite{Poole}. The IBFM can be
considered a model for the normal state, and its temperature instability a
prediction of $T_{c}$.

9. Phase diagram \cite{PLA2} with 3D superconducting critical temperature
phase boundaries $T_{cs+}$, $T_{cs-}$, $T_{css+}$, and $T_{css-}$ as
functions of $\Delta n\equiv n/n_{f}-1$ as defined in text,\ in the vicinity
of the approximate BCS $T_{c}$ value, assuming the quadratic boson
dispersion $\eta =\hbar ^{2}K^{2}/2(2m)$, for $\lambda =1/5$ and $\hbar
\,\omega _{D}/E_{F}=$ $0.001$.

10. Phase diagram in 2D temperature $T$ (in units of $T_{F}$) and electron
density $n$ (in units of $n_{f}$ as defined in text) showing the phase
boundaries of $T_{c}$'s for pure 2e-CP BEC phases (dashed curves) determined
by $\Delta (T_{c})=f\sqrt{n_{0}(T_{c})}$ $\equiv 0$ and pure 2h-CP BEC
phases (full curves) given by $\Delta (T_{c})=f\sqrt{m_{0}(T_{c})}\equiv 0$
for $\lambda =1/4$ and $1/2$ with $\hbar \omega _{D}/E_{F}=0.05$.
Intersections corresponding to $n_{0}(T)=m_{0}(T)$ approximately reproduce
the BCS $T_{c}$ as given by (\ref{sqrt}) and are marked by black dots. Black
squares mark the BEC limit where all electrons are imagined paired into
2e-CP bosons, as calculated in (\ref{52}), to which values the dashed curves
reach only approximately in the limit $n/n_{f}$ $\rightarrow \infty $.

11. Same as Fig. 10 except that $n/n_{f}$ is eliminated numerically and
abscissa refers to $n/n_{f}(T_{c}),$ where $n_{f}(T_{c})$ is the actual
number of unpaired electrons at $T_{c}$ as given by (\ref{55}). Here, the
intersection corresponding to $n_{0}(T)=m_{0}(T)$ occurs at $T=0$ and 
\textit{not} at the approximate BCS $T_{c}$ values as in Fig. 10. Open
circles mark values of $n/n_{f}(T_{c})$ corresponding to $n/n_{f}=\infty $.

12. Same as Fig. 10 but in 3D, and only for $\lambda ={\frac12}$.

13. Same as Fig. 11 but in 3D, and only for $\lambda ={\frac12}$.

\bigskip

{\Large Appendix A. \ Why CPs are bosons while BCS pairs are not.}

This assertion is simply visualized \textit{qualitatively} with the aid of
Fig. 2 for the BCS model interaction (\ref{13}). The vector $\mathbf{k}$\
ends in all points of a simple-cubic lattice in $k$-space\ with lattice
spacing $2\pi /L$ with $L$ the system size. In 3D these points are within
the overlap volume (shaded in the figure) where the interaction is nonzero.
But, in the thermodynamic limit there are infinitely many acceptable $%
\mathbf{k}$\ values allowed for the energy $\mathcal{E}_{K}$ of the CP
state, whether for ordinary CPs (\ref{5}), (\ref{eq:cooper}) or for
Bethe-Salpeter CPs (\ref{mCP}). This implies BE statistics for either
ordinary or BS CPs as each CP energy $\mathcal{E}_{K}$\ level has no
occupation limit. Not so for BCS pairs as we now show.

More \textit{quantitatively}, consider fermion creation $a_{\mathbf{k_{1}}%
s}^{\dag }$\ and annihilation $a_{\mathbf{k_{1}}s}$\ operators that satisfy
the anti-commutation relations.

\begin{equation}
\begin{array}{ll}
\{a_{\mathbf{k_{1}}s}^{\dag },a_{\mathbf{k_{1}^{\prime }}s^{\prime
}}^{\dagger }\}=\{a_{\mathbf{k_{1}}s},a_{\mathbf{k_{1}^{\prime }}s^{\prime
}}\}=0 &  \\ 
\{a_{\mathbf{k_{1}}s},a_{\mathbf{k_{1}^{\prime }}s^{\prime }}^{\dag
}\}=\delta _{\mathbf{k_{1}k_{1}^{\prime }}}\delta _{ss^{\prime }}. & 
\end{array}
\tag{A.1}  \label{A.1}
\end{equation}%
The BCS pair annihilation and creation operators, respectively, are then%
\begin{equation}
b_{\mathbf{kK}}\equiv a_{\mathbf{k}_{2}\downarrow }a_{\mathbf{k}_{1}\uparrow
}\text{ \ \ \ \ and \ \ \ \ }b_{\mathbf{kK}}^{\dag }\equiv a_{\mathbf{k}%
_{1}\uparrow }^{\dag }a_{\mathbf{k}_{2}\downarrow }^{\dag }.  \tag{A.2}
\label{A.2}
\end{equation}%
Here 
\begin{equation}
\mathbf{k}\equiv {\frac{1}{2}}(\mathbf{k_{1}-k_{2})}\ \text{\ \ \ \ and}\ \
\ \ \ \mathbf{K}\equiv \mathbf{k_{1}}+\mathbf{k_{2}}  \tag{A.3}  \label{A.3}
\end{equation}%
are the relative and total (or center-of-mass) momenta wavevectors,
respectively, associated with two fermions with wavevectors 
\begin{equation}
\mathbf{k}_{1}=\mathbf{K}/2+\mathbf{k}\ \text{\ \ \ \ \ and \ \ \ \ }\mathbf{%
k}_{2}=\mathbf{K}/2-\mathbf{k}.  \tag{A.4}  \label{A.4}
\end{equation}%
We show below that $b_{\mathbf{kK}}$ and $b_{\mathbf{kK}}^{\dag }$ satisfy:
a) the sometimes called ``pseudo-boson'' commutation relations%
\begin{equation}
\lbrack b_{\mathbf{kK}},b_{\mathbf{k}^{\prime }\mathbf{K}}^{\dag }]=(1-n_{%
\mathbf{K}/2-\mathbf{k}\downarrow }-n_{\mathbf{K}/2+\mathbf{k}\uparrow
})\delta _{\mathbf{kk}^{\prime }},  \tag{A.5}  \label{A.5}
\end{equation}

\begin{equation}
\lbrack b_{\mathbf{kK}}^{\dag },b_{\mathbf{k}^{\prime }\mathbf{K}}^{\dag }]=%
\left[ b_{\mathbf{kK}},b_{\mathbf{k}^{\prime }\mathbf{K}}\right] =0, 
\tag{A.6}  \label{A.6}
\end{equation}%
where 
\begin{equation}
n_{\mathbf{K}/2\pm \mathbf{k}\downarrow }\equiv {a_{\mathbf{K}/2\pm \mathbf{k%
}\downarrow }^{\dag }}{a_{\mathbf{K}/2\pm \mathbf{k}\downarrow }}  \tag{A.7}
\label{A.7}
\end{equation}%
are fermion number operators; as well as b) a ``pseudo-fermion''
anti-commutation relation%
\begin{equation}
\left\{ b_{\mathbf{kK}},b_{\mathbf{k}^{\prime }\mathbf{K}}\right\} =2b_{%
\mathbf{kK}}b_{\mathbf{k}^{\prime }\mathbf{K}}(1-\delta _{\mathbf{kk}%
^{\prime }}).  \tag{A.8}  \label{A.8}
\end{equation}%
Our only restriction is that $\mathbf{K\equiv k}_{1}+\mathbf{k}_{2}=\mathbf{k%
}_{1}^{\prime }+\mathbf{k}_{2}^{\prime }$. Clearly, BCS pairs are \textit{%
not }bosons as they do \textit{not} satisfy (Ref. \cite{Schrieffer} p. 38)\
the ordinary boson commutation relations 
\begin{equation}
\lbrack b_{\mathbf{kK}},b_{\mathbf{k}^{\prime }\mathbf{K}}^{\dag }]=\delta _{%
\mathbf{kk}^{\prime }}\text{ \ \ \ \ and \ \ \ \ }[b_{\mathbf{kK}}^{\dag
},b_{\mathbf{k}^{\prime }\mathbf{K}}^{\dag }]=\left[ b_{\mathbf{kK}},b_{%
\mathbf{k}^{\prime }\mathbf{K}}\right] =0.  \tag{A.9}  \label{A.9}
\end{equation}

If $\mathbf{K}=0$ (so that $\mathbf{k}_{1}=-\mathbf{k}_{2}=\mathbf{k}$), and
calling $b_{\mathbf{kK=0}}\equiv b_{\mathbf{k}}$, etc., (\ref{A.5}) and (\ref%
{A.6}) become Eqs. (2.11) and (2.12) of Ref. \cite{bcs}, and (\ref{A.8})
becomes Eq. (2.13) thereof.

To prove (\ref{A.5}) to (\ref{A.8}) write 
\begin{equation}
\lbrack {b_{\mathbf{kK}},b_{\mathbf{k}^{\prime }\mathbf{K}}^{\dag }]\equiv
b_{\mathbf{kK}}b_{\mathbf{k}^{\prime }\mathbf{K}}^{\dag }-b_{\mathbf{k}%
^{\prime }\mathbf{K}}^{\dag }b_{\mathbf{kK}}}\equiv a_{\mathbf{k}%
_{2}\downarrow }a_{\mathbf{k}_{1}\uparrow }a_{\mathbf{k}_{1}^{\prime
}\uparrow }^{\dag }a_{\mathbf{k}_{2}^{\prime }\downarrow }^{\dag }-a_{%
\mathbf{k}_{1}^{\prime }\uparrow }^{\dag }a_{\mathbf{k}_{2}^{\prime
}\downarrow }^{\dag }a_{\mathbf{k}_{2}\downarrow }a_{\mathbf{k}_{1}\uparrow }
\tag{A.10}  \label{A.10}
\end{equation}%
using (\ref{A.2}). This can alternately be rewritten, using (\ref{A.4}), as 
\begin{equation}
\lbrack {b_{\mathbf{kK}},b_{\mathbf{k}^{\prime }\mathbf{K}}^{\dag }]=a_{%
\mathbf{K}/2-\mathbf{k}\downarrow }}{a_{\mathbf{K}/2+\mathbf{k}\uparrow }}{%
a_{\mathbf{K}/2+\mathbf{k}^{\prime }\uparrow }^{\dag }}{a_{\mathbf{K}/2-%
\mathbf{k}^{\prime }\downarrow }^{\dag }}-{a_{\mathbf{K}/2+\mathbf{k}%
^{\prime }\uparrow }^{\dag }}{a_{\mathbf{K}/2-\mathbf{k}^{\prime }\downarrow
}^{\dag }}{a_{\mathbf{K}/2-\mathbf{k}\downarrow }}{a_{\mathbf{K}/2+\mathbf{k}%
\uparrow }}.  \tag{A.11}  \label{A.11}
\end{equation}%
The pair of Fermi operators ${a_{\mathbf{K}/2+\mathbf{k}\uparrow }}{a_{%
\mathbf{K}/2+\mathbf{k}^{\prime }\uparrow }^{\dag }}$ in the first term
gives, using (\ref{A.1}),

\begin{equation*}
{a_{\mathbf{K}/2-\mathbf{k}\downarrow }}(\delta _{\mathbf{kk}^{\prime }}-{a_{%
\mathbf{K}/2+\mathbf{k}^{\prime }\uparrow }^{\dag }}{a_{\mathbf{K}/2+\mathbf{%
k}\uparrow }}){a_{\mathbf{K}/2-\mathbf{k}^{\prime }\downarrow }^{\dag }}
\end{equation*}%
\begin{equation}
={a_{\mathbf{K}/2-\mathbf{k}\downarrow }}{a_{\mathbf{K}/2-\mathbf{k}^{\prime
}\downarrow }^{\dag }}\delta _{\mathbf{kk}^{\prime }}-{a_{\mathbf{K}/2-%
\mathbf{k}\downarrow }}{a_{\mathbf{K}/2+\mathbf{k}^{\prime }\uparrow }^{\dag
}}{a_{\mathbf{K}/2+\mathbf{k}\uparrow }}{a_{\mathbf{K}/2-\mathbf{k}^{\prime
}\downarrow }^{\dag }.}  \tag{A.12}  \label{A.12}
\end{equation}%
Using (\ref{A.1}) again the first term becomes $\delta _{\mathbf{kk}^{\prime
}}(1-{a_{\mathbf{K}/2-\mathbf{k}^{\prime }\downarrow }^{\dag }a_{\mathbf{K}%
/2-\mathbf{k}\downarrow }}).$ The last term of (\ref{A.12}), after
anticommuting all creation operators to the left and recalling that $\delta
_{\uparrow \downarrow }\equiv 0$, etc., gives $-\delta _{\mathbf{kk}^{\prime
}}{a_{\mathbf{K}/2+\mathbf{k}^{\prime }\uparrow }^{\dag }}{a_{\mathbf{K}/2+%
\mathbf{k}\uparrow }}$ $+$ ${a_{\mathbf{K}/2+\mathbf{k}^{\prime }\uparrow
}^{\dag }}{a_{\mathbf{K}/2-\mathbf{k}^{\prime }\downarrow }^{\dag }}{a_{%
\mathbf{K}/2-\mathbf{k}\downarrow }}{a_{\mathbf{K}/2+\mathbf{k}\uparrow }}$.
Inserting this in (\ref{A.11}) leaves precisely (\ref{A.5}) if we recall the
number operators (\ref{A.7}). To prove (\ref{A.6}) write%
\begin{equation}
{\left[ b_{\mathbf{kK}},b_{\mathbf{k}^{\prime }\mathbf{K}}\right] \equiv b_{%
\mathbf{kK}}b_{\mathbf{k}^{\prime }\mathbf{K}}-b_{\mathbf{k}^{\prime }%
\mathbf{K}}b_{\mathbf{kK}}}\equiv a_{\mathbf{k}_{2}\downarrow }a_{\mathbf{k}%
_{1}\uparrow }a_{\mathbf{k}_{2}^{\prime }\downarrow }a_{\mathbf{k}%
_{1}^{\prime }\uparrow }-a_{\mathbf{k}_{2}^{\prime }\downarrow }a_{\mathbf{k}%
_{1}^{\prime }\uparrow }a_{\mathbf{k}_{2}\downarrow }a_{\mathbf{k}%
_{1}\uparrow }  \tag{A.13}  \label{A.13}
\end{equation}%
using (\ref{A.2}); or equivalently 
\begin{equation}
{\left[ b_{\mathbf{kK}},b_{\mathbf{k}^{\prime }\mathbf{K}}\right] }\equiv a_{%
\mathbf{K}/2-{\mathbf{k}\downarrow }}a_{\mathbf{K}/2+{\mathbf{k}\uparrow }%
}a_{\mathbf{K}/2-\mathbf{k}^{\prime }{\downarrow }}a_{\mathbf{K}/2+\mathbf{k}%
^{\prime }{\uparrow }}-a_{\mathbf{K}/2-\mathbf{k}^{\prime }{\downarrow }}a_{%
\mathbf{K}/2+\mathbf{k}^{\prime }{\uparrow }}a_{\mathbf{K}/2-{\mathbf{k}%
\downarrow }}a_{\mathbf{K}/2+{\mathbf{k}\uparrow .}}\,  \tag{A.14}
\label{A.14}
\end{equation}%
The first term is easily brought into a form cancelling the last term by
simply anticommuting operators with primed subindices to the left, thus
proving (\ref{A.6}). Finally to prove (\ref{A.8}) write\newline
\begin{equation}
\{{b_{\mathbf{kK}},b_{\mathbf{k}^{\prime }\mathbf{K}}\}\equiv b_{\mathbf{kK}%
}b_{\mathbf{k}^{\prime }\mathbf{K}}+b_{\mathbf{k}^{\prime }\mathbf{K}}b_{%
\mathbf{kK}}}\equiv a_{\mathbf{k}_{2}\downarrow }a_{\mathbf{k}_{1}\uparrow
}a_{\mathbf{k}_{2}^{\prime }\downarrow }a_{\mathbf{k}_{1}^{\prime }\uparrow
}+a_{\mathbf{k}_{2}^{\prime }\downarrow }a_{\mathbf{k}_{1}^{\prime }\uparrow
}a_{\mathbf{k}_{2}\downarrow }a_{\mathbf{k}_{1}\uparrow }  \tag{A.15}
\label{A.15}
\end{equation}%
\begin{equation}
\equiv a_{\mathbf{K}/2-{\mathbf{k}\downarrow }}a_{\mathbf{K}/2+{\mathbf{k}%
\uparrow }}a_{\mathbf{K}/2-\mathbf{k}^{\prime }{\downarrow }}a_{\mathbf{K}/2+%
\mathbf{k}^{\prime }\uparrow }+a_{\mathbf{K}/2-\mathbf{k}^{\prime }{%
\downarrow }}a_{\mathbf{K}/2+\mathbf{k}^{\prime }{\uparrow }}a_{\mathbf{K}/2-%
{\mathbf{k}\downarrow }}a_{\mathbf{K}/2+{\mathbf{k}\uparrow }},  \tag{A.16}
\label{A.16}
\end{equation}%
\begin{equation}
=2a_{\mathbf{K}/2-\mathbf{k}{\downarrow }}a_{\mathbf{K}/2+\mathbf{k}{%
\uparrow }}a_{\mathbf{K}/2-{\mathbf{k}}^{\prime }{\downarrow }}a_{\mathbf{K}%
/2+{\mathbf{k}}^{\prime }{\uparrow }}\equiv 2b_{\mathbf{kK}}b_{\mathbf{k}%
^{\prime }\mathbf{K}}\,\ \ \ \ \ \ \ \ \,\,\,\,\,\,\,\,\,\,\,\,\,\text{if}%
\,\,\,\ \mathbf{k}\neq \mathbf{k}^{\prime }.  \tag{A.17}  \label{A.17}
\end{equation}%
However, if $\mathbf{k}=\mathbf{k}^{\prime }$ then%
\begin{equation}
\{b_{\mathbf{kK}},b_{\mathbf{kK}}\}=2a_{\mathbf{K}/2-\mathbf{k}{\downarrow }%
}a_{\mathbf{K}/2+\mathbf{k}\uparrow }a_{\mathbf{K}/2-{\mathbf{k}\downarrow }%
}a_{\mathbf{K}/2+{\mathbf{k}\uparrow }}=0  \tag{A.18}  \label{A.18}
\end{equation}%
since $(a_{\mathbf{K}/2\pm {\mathbf{k}\downarrow }})^{2}\equiv 0$. Hence (%
\ref{A.8}) is true.

The main point here is simply this: \textit{any} number of CPs with a
definite $\mathbf{K}$ but \textit{different }values of $\mathbf{k}$ can
occupy a state of CP energy $\mathcal{E}_{K}$ and thus not only obey BE
statistics but (as this in turn demands) also obey the boson commutation
relations (\ref{A.5}) for $\mathbf{k\neq k}^{\prime }$. Hence, CPs can
suffer a BEC as a given CP state involves no two BCS pairs with the same $%
\mathbf{k}$.{}

\bigskip

{\Large Appendix B. \ CBFM and BCS condensation energies compared.}

Using (\ref{29}) for $T=0$ when $n_{0}(T)=m_{0}(T)$ and $n_{B}(T)=m_{B}(T)$
we obtain for the superconducting state, calling $\Delta (T=0)\equiv \Delta ,
$%
\begin{equation*}
\frac{\Omega _{s}(T=0)}{L^{d}}=2\hbar \omega _{D}n_{0}(0)+\int_{-\mu
}^{\infty }d\xi N(\xi )[\xi -\sqrt{\xi ^{2}+\Delta ^{2}}]
\end{equation*}%
where $\xi \equiv \varepsilon -\mu $. Recalling the expression (\ref{34})
for $\Delta $, using (\ref{f+}) and (\ref{f-}), and since $\mu =E_{f}$,
gives 
\begin{equation}
\frac{\Omega _{s}(T=0)}{L^{d}}=2\hbar \omega _{D}n_{0}(0)+2\int_{-\mu
}^{-\hbar \omega _{D}}d\xi N(\xi )\xi +N(E_{F})\int_{-\hbar \omega
_{D}}^{\hbar \omega _{D}}d\xi \lbrack \xi -\sqrt{\xi ^{2}+\Delta ^{2}}]. 
\tag{B.1}  \label{B.1}
\end{equation}%
The first and second members of the last term have respectively odd and even
integrands, so that this term becomes 
\begin{equation*}
-2N(E_{F})\int_{0}^{\hbar \omega _{D}}d\xi \lbrack \sqrt{\xi ^{2}+\Delta ^{2}%
}].
\end{equation*}%
On the other hand, for the normal state $n_{0}=0$ and $m_{0}=0$, so that
from (\ref{34}) $\Delta (T)=0$. Hence, from (\ref{45a}) the CBFM
condensation energy $E_{s}-E_{n\text{ }}$per unit volume\ is just%
\begin{equation}
\frac{E_{s}-E_{n}}{L^{d}}=2\hbar \omega
_{D}n_{0}(0)+2N(E_{F})\int_{0}^{\hbar \omega _{D}}d\xi (\xi -\sqrt{\xi
^{2}+\Delta ^{2}})\hspace{1cm}\hbox{(CBFM)}.  \tag{B.2}  \label{B.2}
\end{equation}%
>From Eq. (2), Sec. 4.3.3.1, Ref. \cite{Jeffrey} the integral evaluates to 
\begin{eqnarray*}
&&-\frac{1}{2}\hbar \omega _{D}\sqrt{(%
h{\hskip-.2em}\llap{\protect\rule[1.1ex]{.325em}{.1ex}}{\hskip.2em}%
\omega _{D})^{2}+\Delta ^{2}}+\frac{(\hbar \omega _{D})^{2}}{2}+\frac{1}{2}%
\Delta ^{2}\ln \frac{\Delta }{\hbar \omega _{D}+\sqrt{(\hbar \omega
_{D})^{2}+\Delta ^{2}}} \\
&=&\frac{1}{2}\Delta ^{2}\ln (\Delta /2\hbar \omega _{D})-\frac{1}{4}\Delta
^{2}-\frac{1}{16}[\Delta ^{4}/(\hbar \omega _{D})^{2}]+O\left[ \Delta
^{6}/(\hbar \omega _{D})^{4}\right] .
\end{eqnarray*}%
Thus, since $n_{0}=\Delta ^{2}/f^{2}$, (\ref{B.2}) for weak coupling $\Delta
=f\sqrt{n_{0}}\rightarrow 0$ becomes 
\begin{equation}
\frac{E_{s}-E_{n}}{L^{d}}\smash
{\mathop{\relbar\joinrel\longrightarrow}\limits_{\lambda \to 0}}-\frac{1}{2}%
N(E_{F})\Delta ^{2}\left[ 1+\frac{1}{4}\left( \Delta /\hbar \omega
_{D}\right) ^{2}+O\left( \Delta /\hbar \omega _{D}\right) ^{4}\right] 
\hspace{1cm}\hbox{(CBFM)}.  \tag{B.3}  \label{B.3}
\end{equation}

By contrast, the original expressions (2.41) and (2.43) in Ref. \cite{bcs}
give 
\begin{equation}
\frac{E_{s}-E_{n}}{L^{d}}=2N(E_{F})\int_{0}^{\hbar \omega _{D}}d\xi \left(
\xi -\frac{\xi ^{2}}{\sqrt{\xi ^{2}+\Delta ^{2}}}\right) -\frac{\Delta ^{2}}{%
V}\hspace{1cm}\hbox{(BCS)}  \tag{B.4}  \label{B.4}
\end{equation}%
where $V$ is defined in (\ref{13}). Multiplying (\ref{20}) by $\Delta ^{2}/2$
is equivalent to 
\begin{equation*}
2N(E_{F})\int_{0}^{\hbar \omega _{D}}d\xi \frac{\Delta ^{2}}{2\sqrt{\xi
^{2}+\Delta ^{2}}}=\frac{\Delta ^{2}}{V}
\end{equation*}%
which when combined with (\ref{B.4}) gives 
\begin{equation*}
\frac{E_{s}-E_{n}}{L^{d}}=2N(E_{\mbox{\tiny F}})\int_{0}^{\hbar \omega _{%
\mbox{\tiny D}}}d\xi \bigg[\xi -{\frac{1}{2}}{\frac{2\xi ^{2}+\Delta ^{2}}{%
\sqrt{\xi ^{2}+\Delta ^{2}}}}\bigg].
\end{equation*}%
Using Eqs. (3) and (10), Sec. 4.3.3.1, of Ref. \cite{Jeffrey} finally gives 
\begin{equation}
\frac{E_{s}-E_{n}}{L^{d}}=N(E_{F})(\hbar \omega _{D})^{2}\left[ 1-\sqrt{%
1+\left( \Delta /\hbar \omega _{D}\right) ^{2}}\right]   \tag{B.5}
\label{B.5}
\end{equation}%
which on expansion leads to 
\begin{equation}
\frac{E_{s}-E_{n}}{L^{d}}\smash
{\mathop{\relbar\joinrel\longrightarrow}\limits_{\lambda \to 0}}-\frac{1}{2}%
N(E_{F})\Delta ^{2}\left[ 1-\frac{1}{4}\left( \Delta /\hbar \omega
_{D}\right) ^{2}+O(\Delta /\hbar \omega _{D})^{4}\right] \hspace{1cm}%
\hbox{(BCS)}.  \tag{B.6}  \label{B.6}
\end{equation}%
\qquad 

Thus, the CBFM condensation energy (\ref{B.3})\ is \textit{larger }than that
of BCS, but only the latter is a rigorous upper bound to the exact
condensation energy since it follows \cite{bcs} from a variational trial
wavefunction.

\bigskip

{\Large Appendix C. \ BEC as limit of all electrons paired.}

Here we discuss generalizations of the well-known result that the BEC
transition temperature $T_{c}$ of a $d$-dimensional $N$-fermion gas of Fermi
temperature $T_{F}$ in which \textit{all }fermions are imagined paired into
bosons, is just $0.218T_{F}$. (See dashed line in ``Uemura plot'' of Ref. %
\cite{Uemura}, Fig. 2). These results will provide a convenient pure BEC
limit which in effect turn out to be an \textit{upper limit }for the pure
2e-CP phase separation boundary $T_{c}$ values deduced in Sec. 8.

The general BEC $T_{c}$-formula for identical noninteracting bosons in $d$%
-dimensions of energy $\eta =C_{s}\,K^{s},$ $s>0,$ where as before $K$ is
the boson CMM, is \cite{B8a}

\begin{equation}
T_{c}=\frac{C_{s}}{k_{B}}\left[ \frac{s\Gamma (d/2)(2\pi )^{d}}{2\pi
^{d/2}\Gamma (d/s)g_{d/s}(1)}n_{B}\right] ^{s/d}  \tag{C.1}  \label{C.1}
\end{equation}%
where $n_{B}$ is the boson number-density and $g_{\sigma }(z)$ the Bose
integral \cite{Path}, $z\equiv e^{\mu _{B}/k_{B}T}$ is the ``\textit{fugacity%
}''\ and $\mu _{B}$ the boson chemical potential. For $z=1$, $g_{\sigma }(1)$
$\equiv $ $\zeta (\sigma )$, the Riemann Zeta-function, if $\sigma >1$,
while for $0<\sigma \leq 1$ the infinite series $g_{\sigma }(1)$ diverges.\
Eq. (\ref{C.1}) is formally valid for all $d>0$ and $s>0$. Hence, for $%
0<d\leq s$, $T_{c}=0$ since $g_{d/s}(1)=\infty $ for $d/s\leq 1$ but $T_{c}$
is otherwise finite. We stress that as a consequence of the former \textit{%
all }2D\textit{\ }$T_{c}$ predictions in Fig. 7 (except the BCS one that
survives for all $d>0$, including $d=1$) would collapse to \textit{zero} had 
$s=2$ been used in 2D instead of the correct $s=1$ arising from the Fermi
sea. For $s=2$, $C_{2}=\hbar ^{2}/2m_{B}$ (\ref{C.1})\ leads to the familiar
3D result 
\begin{equation}
T_{c}\simeq {3.31\hbar ^{2}n_{B}^{2/3}/}m_{B}k_{B}  \tag{C.2}  \label{C.2}
\end{equation}%
since $\zeta (3/2)\simeq 2.612$. Recalling that $k_{B}T_{F}=$ $\hbar
^{2}k_{F}^{2}/2m$ with $k_{F}=[2^{d-2}\pi ^{d/2}d\Gamma (d/2)n]^{1/d}$ from (%
\ref{24}), then if $m_{B}=2m$ and $n_{B}=n/2$ (all electrons paired) for $%
s=2 $ (\ref{C.1}) implies that 
\begin{equation}
T_{c}/T_{F}=\tfrac{1}{2}[2/d\Gamma (d/2)g_{d/2}(1)]^{2/d}=0\text{ \ \ \ for
\ \ \ }d\leq 2  \tag{C.3}  \label{C.3}
\end{equation}%
since $g_{d/2}(1)=\infty $ for $d/2\leq 1$. When $d>2$ then $T_{c}$ is
nonzero. For $d=3$ we get the familiar limit mentioned before 
\begin{equation}
T_{c}/T_{F}=\tfrac{1}{2}\left[ 2/3\Gamma (3/2)\zeta (3/2)\right]
^{2/3}\simeq 0.218,  \tag{C.4}  \label{C.4}
\end{equation}%
(see also dashed line Fig. 2 of Ref. \cite{Uemura}).

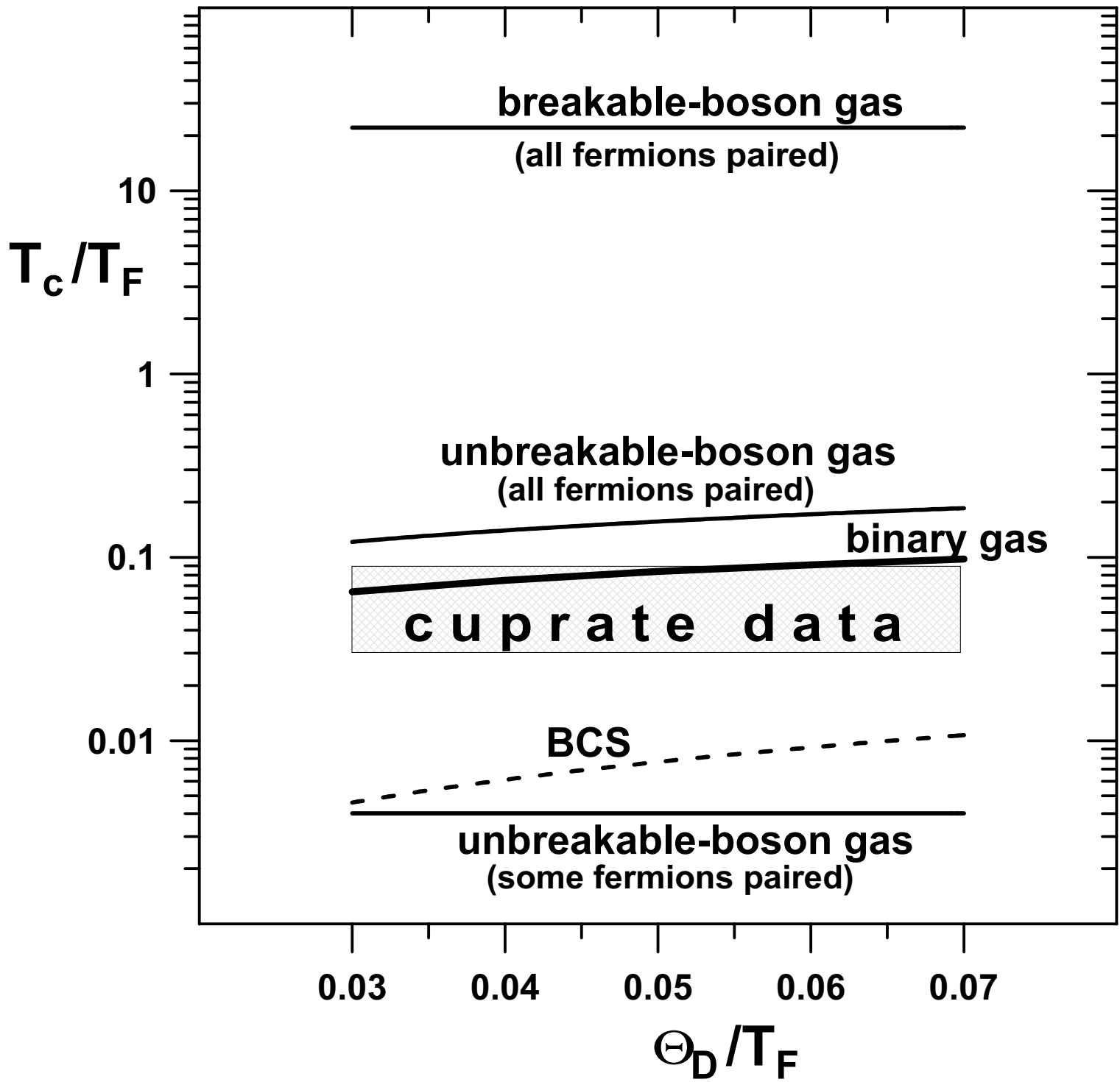
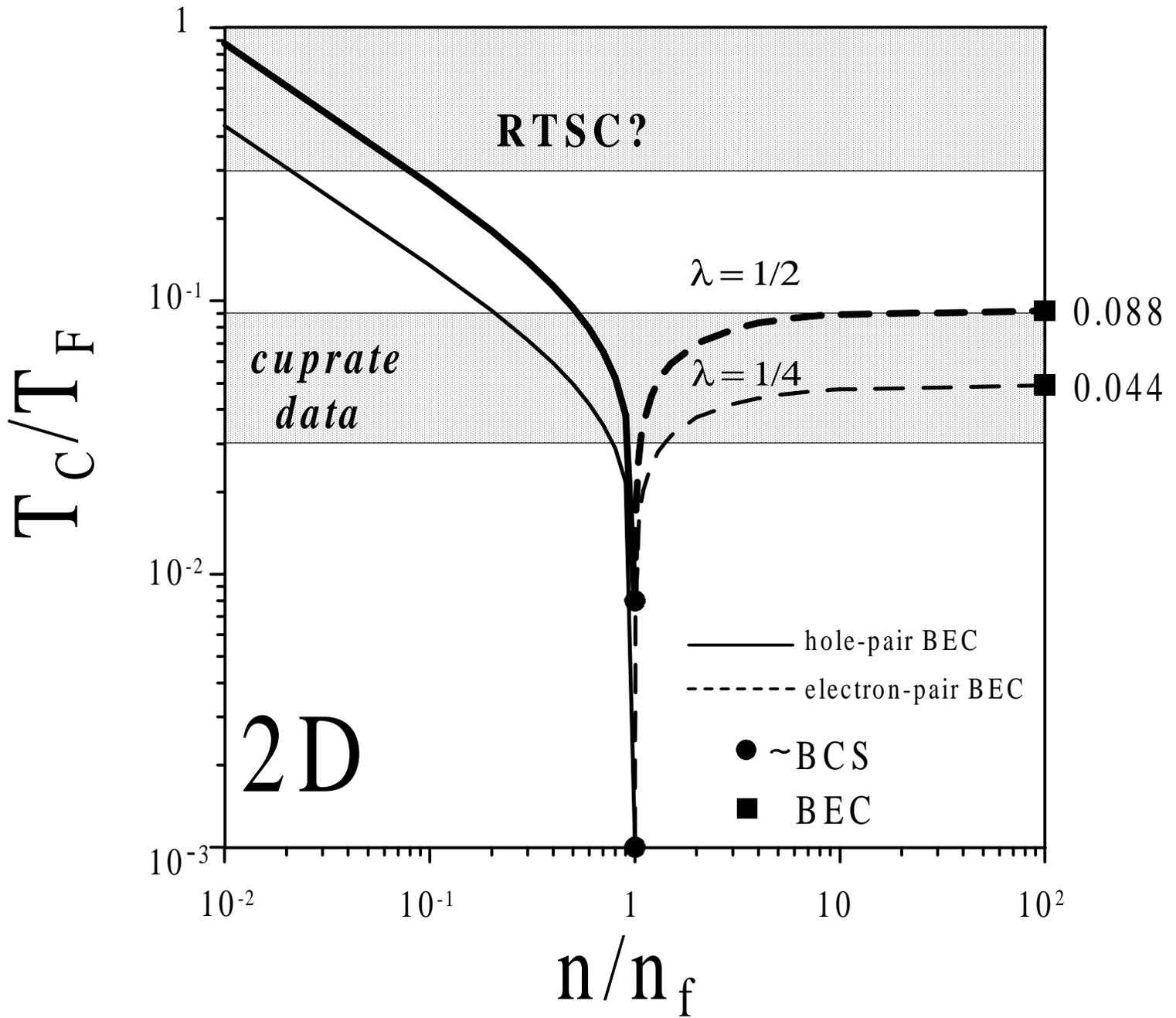
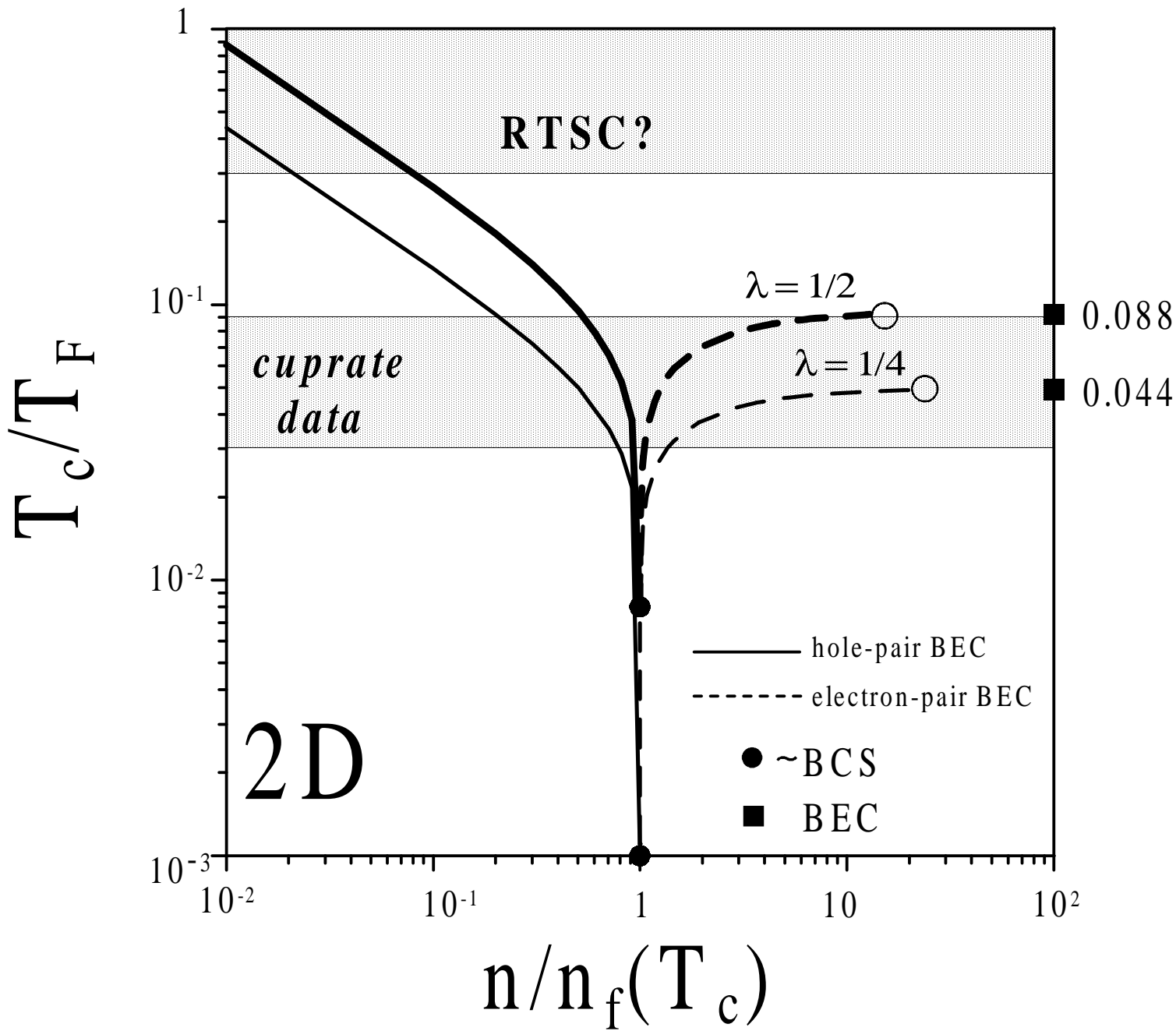
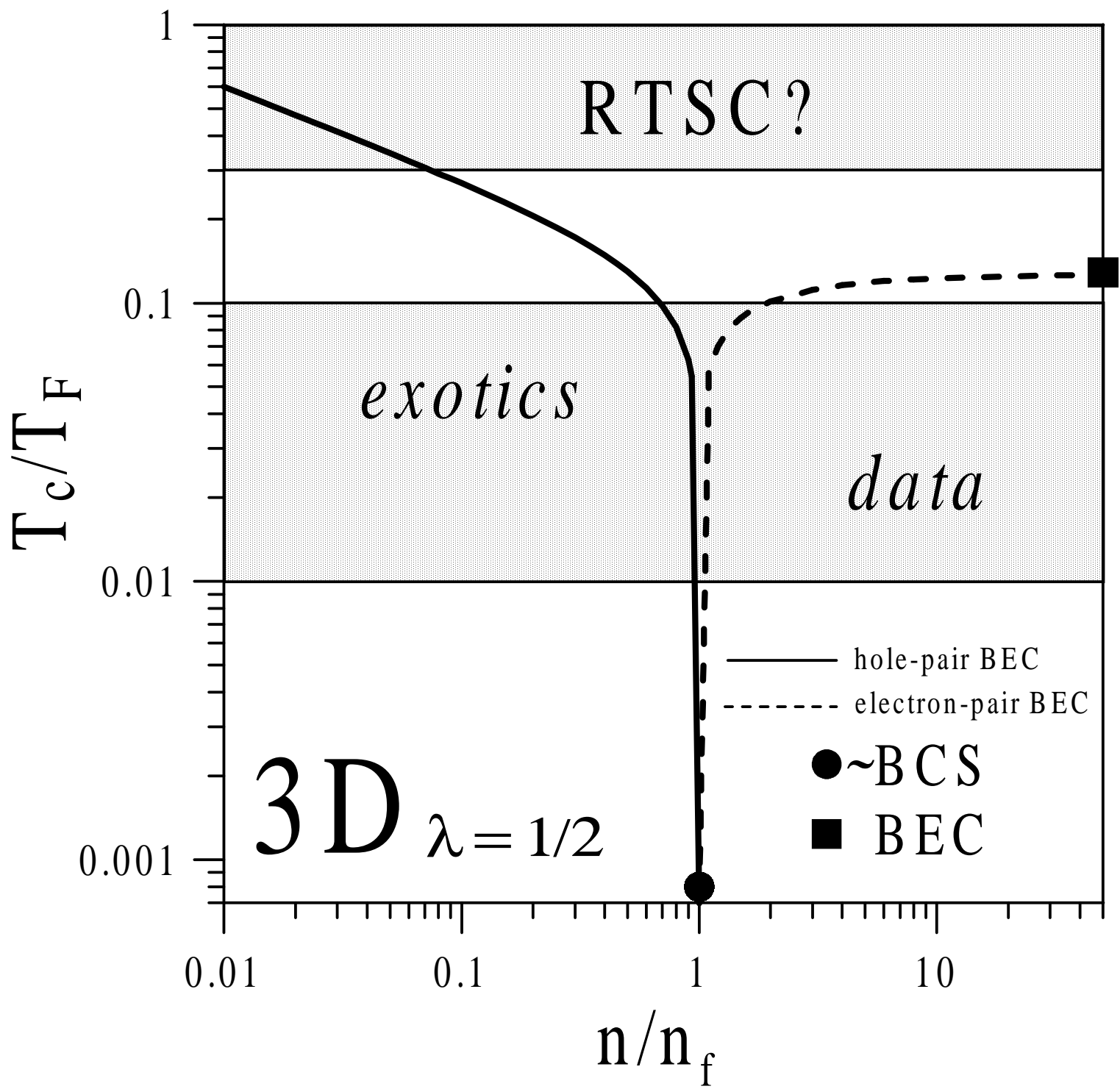
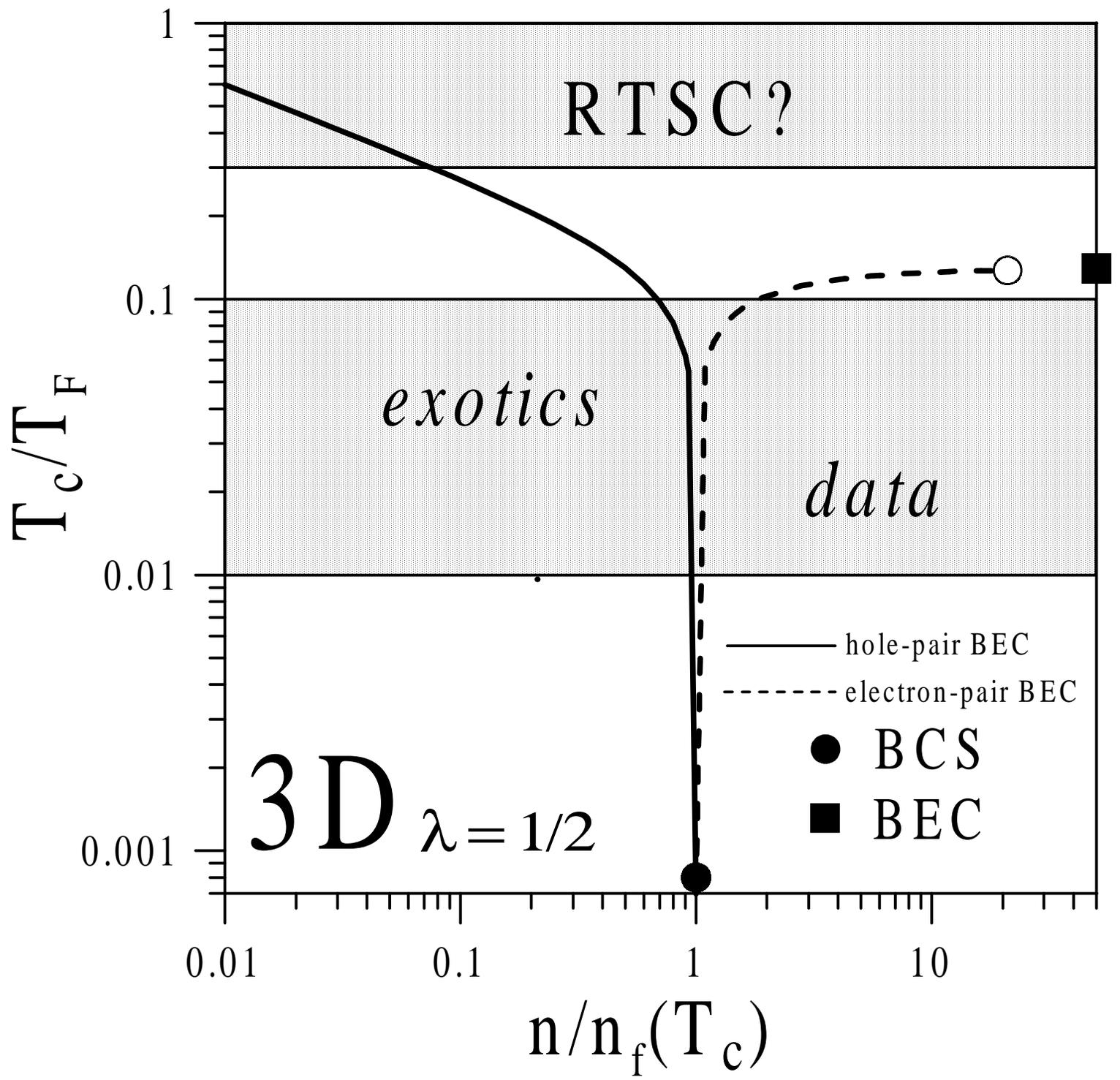
\end{document}